\documentclass[a4paper, 10pt]{article}
\usepackage[latin1]{inputenc}

\pdfoutput=1

\usepackage{jcappub} 
                     
\usepackage{amssymb, amsmath}
\usepackage[below]{placeins}
\usepackage{verbatim}
\usepackage{amsmath, amssymb, amsfonts }
\usepackage{helvet}
\usepackage{footnote}
\usepackage{subfig}
\usepackage{booktabs}

\usepackage{microtype}

\graphicspath{{figures/initial_conditions/},
              {figures/PowerSpectra/},
              {figures/coupling_strength/},
             }

\renewcommand{\d}{\mathrm{d}}
\newcommand{\p}{_{\|}}
\renewcommand{\o}{_{\perp}}
\newcommand{\st}{_{\ast}}
\newcommand{\lm}{^{(\ell m)}}

\newcommand{\lmast}{^{(\ell m)\ast}}
\newcommand{\lmpast}{^{(\ell' m')\ast}}

\newcommand{\overbar}[1]{\mkern 1.5mu\overline{\mkern-1.5mu#1\mkern-1.5mu}\mkern 1.5mu}

\title{ Linear perturbations in spherically symmetric dust cosmologies including a cosmological constant }

\author[a]{Sven Meyer}
\author[a]{Matthias Bartelmann}

\affiliation[a]{Zentrum f\"ur Astronomie der Universit\"at Heidelberg, Institut f\"ur 
Theoretische Astrophysik, Philosophenweg~12, 69120 Heidelberg, Germany}

\emailAdd{ sven.meyer@uni-heidelberg.de }
\emailAdd{ bartelmann@uni-heidelberg.de }

\abstract{

We study the dynamical behaviour of gauge-invariant linear perturbations in spherically symmetric dust cosmologies including a cosmological constant. In contrast to spatially homogeneous FLRW models, the reduced degree of spatial symmetry causes a non-trivial dynamical coupling of gauge-invariant quantities already at first order perturbation theory and the strength and influence of this coupling on the spacetime evolution is investigated here. We present results on the underlying dynamical equations augmented by a cosmological constant and integrate them numerically. We also present a method to derive cosmologically relevant initial variables for this setup. Estimates of angular power spectra for each metric variable are computed and evaluated on the central observer's past null cone. By comparing the full evolution to the freely evolved initial profiles, the coupling strength will be determined for a best fit radially inhomogeneous patch obtained in previous works (see \cite{redlich_probing_2014}). We find that coupling effects are not noticeable within the cosmic variance limit and can therefore safely be neglected for a relevant cosmological scenario. On the contrary, we find very strong coupling effects in a best fit spherical void model matching the distance redshift relation of SNe which is in accordance with previous findings using parametric void models.             

}


\begin{document}

\maketitle

\section{Introduction}
\label{sect:intro}

The modern standard model of cosmology is based on General Relativity (GR) and two symmetry assumptions stating that (1) the universe is, on average, spatially isotropic around our position and (2) this position in the universe is not distinct. These assumptions allow to construct the generic class of Friedmann-Lema\^itre-Robertson-Walker (FLRW) models that are particularly simple and highly symmetric cosmological solutions. It is absolutely remarkable that these solutions do not only generally agree with observational data, but specific models can even be singled out. Recent observations of type Ia supernovae, the cosmic microwave background, gravitational lensing and large scale structure support the spatially flat $\Lambda$CDM model that describes the observable universe extremely well on a wide range of times and spatial scales (see \cite{bartelmann_dark_2010} for a review). \bigskip

Depite this success, the basic assumptions of these models have to be tested extensively. This work aims at a test of the Copernican Principle as we want to restrict ourselves to models with GR  which is well confirmed as underlying theory of gravity on small and intermediate scales. The strong uniformity of the observed cosmic microwave background (CMB) signal strongly supports the concept of spatial isotropy at least around our position. The Copernican Principle and the resulting spatial homogeneity of our universe on sufficiently large scales is, in fact, difficult to test. Among several possibilities, the construction and analysis of exact spatially inhomogeneous solutions of Einstein's equations has been a useful tool which includes the ($\Lambda$)-Lema\^{i}tre-Tolman-Bondi ( ($\Lambda$)LTB ) solutions (see original works in \cite{lemaitre_univers_1933}, \cite{tolman_effect_1934}, and \cite{bondi_spherically_1947} and \cite{bolejko_inhomogeneous_2011, clarkson_establishing_2012, enqvist_lemaitre_2008, marra_observational_2011} for detailed reviews). These are spherically symmetric dust solutions which contain the dust FLRW model in the limit of spatial homogeneity. As such, they are the simplest possible inhomogeneous generalisation of FLRW models based on an exact solution of GR. \bigskip

Spherical void models based on the LTB solution have extensively been tested and confronted with multiple observational probes. The basic idea has been to create a large spatial variation that could potentially model effects of a late time accelerated expansion in observational data retrieved on the past null cone. While measurements on the local Hubble rate, SNae (\cite{redlich_probing_2014}, \cite{february_rendering_2010}, \cite{clarkson_establishing_2012}) as well as CMB measurements alone (see \cite{clifton_what_2009}) can sufficiently be described by suitable void density and curvature profiles, their ability to match a full combined set of observables is very poor (see \cite{redlich_probing_2014}, \cite{biswas_testing_2010}, \cite{bolejko_testing_2009}, \cite{celerier_we_2000}, \cite{garcia-bellido_confronting_2008}, \cite{moss_precision_2011}, \cite{valkenburg_testing_2014}, \cite{zibin_can_2008}, \cite{zumalacarregui_tension_2012}). In addition, the large intrinsic shear causes a strong anisotropic expansion behaviour away from the void's center. At first, this constrains our position to be extremely close ($\sim$ Mpc) to the void's center (see \cite{alnes_cmb_2006, foreman_spatial_2010} for corresponding CMB analyses) and creates a large kinetic Sunyeaev Zel'dovich signal which by far exceeds current upper bounds obtained from measurements (\cite{bull_kinematic_2012}). We can therefore assume spherical void models based on LTB solutions to be ruled out by observations. \bigskip

Consequently, a non-vanishing cosmological constant seems unevitable in order to accurately and consistently describe multiple cosmological probes. It turns out that spherically symmetric dust solutions of Einstein's field equations can easily be augmented by a cosmological constant yielding to so-called $\Lambda$LTB models that take $\Lambda$ for an additional fit parameter. Those models are an effective tool to study deviations from spatial homogeneity and therefore allow to test the Copernican Principle. Whereas late time accelerated expansion can then be covered globally by the cosmological constant, the radial profile of the local Gpc-scale universe around our position can be modeled. $\Lambda$LTB models have been confronted with combined cosmological observables as well (see \cite{redlich_probing_2014}, \cite{marra_observational_2011}) and remarkably only small \%-level deviations from spatial homogeneity have been found. However, the error bars on these estimates are still quite large such that deviations from spatial homogeneity are not significant. \bigskip

Within very few exceptions (see \cite{zibin_can_2008}, \cite{dunsby_how_2010}), observables covered so far in the multi-probe analyses for $\Lambda$LTB and LTB models are not including any information of the late time evolution on linear perturbations in these models, since linear structure formation in radially inhomogeneous models is substantially more complicated (and therefore less feasible) than in FLRW models. Due to the reduced spatial symmetry, linear structure formation depends on the radial position and perturbations evolve anisotropically when placed away from the center of the inhomogeneous patch. This causes linear scalar-vector-tensor variables to couple dynamically which is described by a coupled system of partial differential equations challenging the numerical treatment. The evolution equations of gauge-invariant linear perturbations in generic spherically symmetric solutions has first been derived by Gerlach \& Sengupta (1978) (\cite{gerlach_homogeneous_1978}, \cite{gerlach_relativistic_1978}, \cite{gerlach_gauge-invariant_1979}, \cite{seidel_gravitational_1990}, \cite{tomita_gauge-invariant_2010}) using a 2+2 split of the full spacetime and spherical harmonic decomposition into polar and axial modes. This has subsequently been brought into a numerically feasible form by Gundlach and Mart\'{i}n-Garc\'{i}a (2000) (\cite{gundlach_gauge-invariant_2000, martin-garcia_gauge-invariant_2001}) using fluid-comoving observers. Clarkson et al. (2009) (\cite{clarkson_perturbation_2009}) then adapted those equations to LTB dust solutions and carefully derived the FLRW limit of the proposed set of gauge-invariant variables and their evolution equations. First numerical investigations have been performed in (\cite{february_evolution_2014}). In case of Gaussian shaped void profiles, coupling effects have for the first time been studied in cosmologically relevant initial conditions in (\cite{meyer_evolution_2015}). It should be mentioned that different approaches exist based on conserved quantities (see \cite{leithes_conserved_2015}) or on a covariant 1+1+2 split of the full spacetime (see \cite{dunsby_how_2010}, \cite{zibin_scalar_2008}) which are not considered for this analysis. Although being very successful in the silent approximation neglecting dynamical coupling, the full dynamical equations turn into highly complicated sets of covariant expressions (see \cite{dunsby_how_2010}, \cite{clarkson_covariant_2007}). We therefore decided to rely on the 2+2 split originally developed by Gerlach \& Sengupta. \bigskip 

We adapt the evolution equations derived in (\cite{clarkson_perturbation_2009}) to $\Lambda$LTB models and use a numerical scheme developed in (\cite{meyer_evolution_2015}) to evolve the polar master equations forward in time. We provide realistic initial conditions based on an initial scalar gravitational potential being decomposed into spherical harmonic modes. Coupling effects are then estimated by comparing the freely evolved with the fully coupled solution. The formalism is applied to the best fit $\Lambda$LTB model constrained by several observational probes in (\cite{redlich_probing_2014}). We are particularly interested in the strength and influence of coupling effects in this cosmologically relevant $\Lambda$LTB model and in a confirmation that those can safely be neglected. Due to the small deviation of a radial density profile of a $\Lambda$LTB patch from a spatially homogeneous $\Lambda$CDM model, those effects are expected to be small. However, due to high complexity of linear perturbation theory in radially inhomogeneous models, this is a priori not clear and needs to be proven very carefully. The work presented here is therefore planned as an intermediate step towards an extension of the analysis on $\Lambda$LTB models including observables from intermediate to late time linear structure formation. \bigskip

The paper is structured as follows. The construction and implementation of the background $\Lambda$LTB solution are laid out Sect. (\ref{sect:background}) followed by expressions for the full set of evolution equations of gauge-invariant linear perturbations in Sect. (\ref{sect:perturbation}). Sects. (\ref{sect:initial}) and (\ref{sect:numeric}) outline the construction of initial and boundary conditions as well as the key aspects of the numerical implementation. Final results on the angular power spectra and coupling strengths are presented and discussed in the final Section (\ref{sect:results}).

\section{Dynamics of \texorpdfstring{$\Lambda$}\ LTB models }
\label{sect:background}

The $\Lambda$LTB solution is a dust solution of Einstein's field equations that contains spatial hypersurfaces being spherically symmetric about a distinct ``central" worldline. The geometrical properties are identical to the LTB solution, but its dynamics are altered by a non-vanishing cosmological constant $\Lambda$.  \bigskip   

As dust worldlines are geodesics this allows to introduce freely falling, comoving observers that are described by comoving synchronous coordinates (see \cite{straumann_general_2013}). In these coordinates the line element reads

\begin{equation}
 \d s^2 = -\d t^2 + \frac{a\p^2(t,r)}{1-\kappa(r) r^2} \d r^2 + r^2 a\o^2(t,r) \d \Omega^2\,, 
 \label{lltb:background:1}
 \end{equation}

and the energy momentum tensor describes a pressureless dust fluid $T_{\mu\nu} = \rho(t,r) u_\mu u_\nu$. The metric defined in Eq. (\ref{lltb:background:1}) is constructed to be asymptotically embedded into a background $\Lambda$CDM model of given background parameters $H_0$, $\Omega_m$ and $\Omega_\Lambda$. As a pure dust solution, the $\Lambda$LTB spacetime does not contain fluids with pressure support like radiation such that we can only treat it as a small and subdominant test field. However, this is considered to be well fulfilled in the redshift range considered in this work.  

Analogously to the LTB case, we define the Hubble rates

\begin{equation}
 H\o(t,r) = \frac{\dot{a}\o(t,r)}{a\o(t,r)}, \ H\p(t,r) = \frac{\dot{a}\p(t,r)}{a\p(t,r)}\,. 
 \label{lltb:background:2}
\end{equation}

Einstein's field equations $G_{\mu\nu} + \Lambda g_{\mu\nu} = 8\pi G \, T_{\mu\nu}$ can then be reduced to two remaining expressions\footnote{Throughout this work, we apply the notation Clarkson (2012) (\cite{clarkson_establishing_2012}) using  $\{ a\o, a\p, M, \kappa\}$. In the context of linear perturbation theory, the similarity of this notation to the familiar FLRW background quantities turns out to be more appropriate then the standard notation $\{ R, R', \tilde{M}, E\}$ that has widely been applied in the literature. For better comparison, these quantities are related via $a\o(t,r) = R(t,r)/r$, $a\p(t,r) = R'(t,r)$, $M(r) = 2\tilde{M}(r)/r^3$ and $\kappa(r) = -2E(r)/r^2$. } 

\begin{align}
  \label{lltb:background:3}
  \frac{(r^3 M(r))'}{r^2 a\o^2 a\p} = 8\pi \rho\,, \\
  \label{lltb:background:4}
  H\o^2 = \frac{M(r)}{a\o^3} - \frac{\kappa(r)}{a\o^2} + \frac{\Lambda}{3}\,.
\end{align}

Eq. (\ref{lltb:background:4}) can be integrated and yields 

\begin{equation}
 t_0 - t_B(r) = \int_0^1 \frac{\sqrt{a\o} \, \d a\o }{\sqrt{(M-\kappa r a\o+\Lambda/3\,r^3 a\o^3)}}\,,
 \label{lltb:background:5}
\end{equation}

with the so-called bang time function as additional degree of freedom. We assume a synchronous big bang by setting $t_B(r) = 0$ for all values of the radial coordinate $r$ in order to avoid decaying modes in a linear approximation of the $\Lambda$LTB patch at early times (see \cite{silk_large-scale_1977}) which would be in contrast to the standard inflationary paradigm. \bigskip 

In comoving synchronous coordinates, the $\Lambda$LTB metric admits a global time coordinate. In particular, the $\Lambda$LTB patch then has the same age as the background $\Lambda$CDM model if we assume a synchronous Big Bang. As done in (\cite{redlich_probing_2014}), we fix the age $t_0$ of the background FLRW universe which is uniquely determined by the background model parameters $H_0$, $\Omega_m$, $\Omega_\Lambda$. We use the well-known gauge freedom in the choice of the areal radius $ra\o(t,r)$ to set $a\o(t_0, r)=1$. Using Eq. (\ref{lltb:background:3}), $M(r)$ then becomes a mass integral given by

\begin{equation}
 M(r) = \frac{8\pi G}{r^3} \int_0^r{\d r' r'^2 \rho(t_0, r')}\,.
 \label{lltb:background:6}
\end{equation}

The density profile at present time can effectively be modelled by a set of nodes $\left\{r_i, \rho(t_0, r_i)\equiv a_i \right\}$ sampling the domain of interest and a corresponding cubic spline interpolation between them (see \cite{redlich_probing_2014} for details). In contrast to the LTB case, there does not exist any parametric solution to the Eq. (\ref{lltb:background:5}), but the resulting elliptic integral can be computed by transforming it to Carlson symmetric forms (see \cite{carlson_numerical_1995}, \cite{valkenburg_complete_2012})

\begin{equation}
 t_0 = \frac{2}{3}\frac{\mathrm{i}}{\sqrt{c}} \frac{1}{(R_1R_2R_3)^{1/2}} R_J \left(\frac{1}{r}-\frac{1}{R_1}, \frac{1}{r}-\frac{1}{R_2}, \frac{1}{r}-\frac{1}{R_3}, \frac{1}{r}\right)\,,
 \label{lltb:background:7}
 \end{equation}

where $c=\Lambda/3$ and $R_i$  are roots of the cubic polynomial $f(R) = M - \kappa R + \frac{\Lambda}{3}R^3$. These forms can be computed very efficiently by a iterative scheme based on certain functional identities. Provided a model for the density profile of the $\Lambda$LTB patch at present time and the parameters of the asymptotic background $\Lambda$CDM model, the mass function $M(r)$ and the global age $t_0$ are fixed such that Eq. (\ref{lltb:background:6}) is a functional of the curvature profile $\kappa(r)$. By using a root finding algorithm, we can determine $\kappa(r)$ numerically. The dynamics of the $\Lambda$LTB patch are then completely determined by evolving Eq. (\ref{lltb:background:4}) in time which yields the scale factor $a\o(t,r)$. In addition, the radial scale factor $a\p(t,r)$ can be expressed by Carlson symmetric forms as well using the orthogonality of coordinate time and radius. As suggested in (\cite{valkenburg_complete_2012}), partial fractioning leads to

\begin{equation}
\begin{split}
  a\p(t,r )= \frac{2 \mathrm{i} \, r\dot{a}\o}{3 c^{3/2}(R_1R_2R_3)^{1/2}} &\left[ \frac{\dfrac{(Mr^3)'}{R_1} - (\kappa r^2)'}{2(R_1-R_2)(R_1-R_3)} R_D\left(\frac{1}{r a\o}-\frac{1}{R_2}, \frac{1}{r a\o}-\frac{1}{R_3}, \frac{1}{r a\o}-\frac{1}{R_1} \right) \right. \\
                                                               &+\left. \text{cyclic permutations in} \ (R_1,R_2,R_3) \frac{}{}\right]\,,
\end{split}
\label{lltb:background:8}
\end{equation}

which is the $\Lambda$LTB generalisation of the well known expression in LTB models

\begin{equation}
 a\p(t,r) = \left(\frac{3}{2} \frac{\kappa'}{\kappa} - \frac{M'}{M}\right) r \dot{a}\o t + \left(\frac{M'}{M} - \frac{\kappa'}{\kappa} + \frac{1}{r}\right) ra\o \,.
\label{lltb:background:9}
\end{equation}

Once the radial scale factor $a\p$ is known in terms of Carlson symmetric forms, $\dot{a}\p$ can easily be computed in a closed form as well:

\begin{equation}
\dot{a}\p(t,r)= \frac{1}{H\o}\left[\frac{3\,M + M'\,r}{2\,a\o^2} - \frac{\kappa + \kappa'\, r}{a\o} + \left( -\frac{M}{2a\o^3} + \frac{\Lambda}{3}\right) a\p \right]\,,
\label{lltb:background:10}
\end{equation}

which fixes the radial Hubble rate $H\p(t,r)$.\bigskip

Throughout this work, we will assume observers located at the center of the $\Lambda$LTB patch. Inward radial null geodesics are then described by the equations 

\begin{align}
 \label{lltb:background:11}
 \frac{\d t(r)}{\d r} &= - \frac{a\p(t(r),r)}{\sqrt{1-\kappa(r) r^2}}\,,\\
 \label{lltb:background:12}
 \frac{1}{ 1 + z(r)} \frac{\d z(r)}{\d r} &=  \frac{\dot{a}\p(t(r),r)}{\sqrt{1-\kappa(r) r^2}}\,,
 \end{align}
 
which fix the central observer's past null cone. Eqs. (\ref{lltb:background:11}) and (\ref{lltb:background:12}) are identical to the LTB case as both spacetimes share the same geometrical properties.

\section{Gauge invariant linear perturbation theory}
\label{sect:perturbation}

Linear perturbation theory in radially inhomogeneous cosmologies is substantially more complicated than in homogeneous and isotropic FLRW models. In the context of spherically symmetric models, Gerlach and Sengupta (see \cite{gerlach_relativistic_1978}) suggested a covariant 2+2 split of the full spacetime ($\mathcal{M}^4 = \mathcal{M}^2 \times \mathcal{S}^2$) which allows to characterise objects in this spacetime according to their transformation properties on the two sphere. In this context, it turns out to be useful to study linear perturbations of spherically symmetric spacetimes in harmonic space by expanding them into scalar, vector and tensor spherical harmonic functions. Perturbations can then naturally be split into a polar (curl-free or even)  and axial (divergence-free or odd) part which are dynamically decoupled. Gundlach \& Mart\'{i}n Garc\'{i}a (GMG) (see \cite{gundlach_gauge-invariant_2000}) adapted this approach to study linear perturbations in the context of stellar collapse which has been specified to spherically symmetric dust spacetimes in Clarkson et al. (2009) (CCF) (see \cite{clarkson_perturbation_2009}). The authors construct a set of gauge-invariant linear perturbations of the LTB spacetime in harmonic space and derive the dynamical equations as well as a rigorous FLRW limit of those which allows a direct comparison of both models. The properties of gauge-invariant linear perturbations in LTB models have extensively been discussed in several papers (\cite{clarkson_perturbation_2009, february_evolution_2014, meyer_evolution_2015}) and, generically, two main complications arise in comparison to spatially homogeneous and isotropic background models:

\begin{itemize}
 \item The more complicated background symmetry causes structure formation to depend on position in the LTB patch. As a result, gauge-invariant linear perturbations do not evolve independently but are dynamically coupled.
 \item Gauge invariant, ``physical", perturbations in LTB spacetimes cannot trivially be mapped to the familiar FLRW scalar-vector-tensor (SVT) variables in the FLRW limit which makes their physical interpretation highly difficult.  
\end{itemize}

The structure of gauge-invariant perturbations in $\Lambda$LTB models is similar to the LTB case as both manifolds have the same geometrical properties. The cosmological constant $\Lambda$ just enters at the background level and does, by construction, not possess any perturbations on its own. Nonetheless, it is a priori not clear if the FLRW limit and the identification of polar and axial modes in terms of SVT modes have the exact same form (especially for the fluid variables). However, repeating the construction of this FLRW limit for the $\Lambda$LTB case yields just trivial differences that do not affect the construction of initial conditions for the cases considered in this work. \bigskip

In fact, we start with the same perturbed metric and energy momentum tensor for the polar branch (see also \cite{february_evolution_2014})

 \begin{align}
  \label{lltb:perturbation:1}
  ds^2 &= -\left[ 1 + (2\eta\lm - \chi\lm - \varphi\lm) Y\lm \right]\d t^2 - \frac{2 a\p \varsigma\lm  Y\lm }{\sqrt{1 - \kappa r^2}} \d t \d r  \\
       & \ \ \ \ + \frac{a\p^2}{1 - \kappa r^2} \left[ 1 + (\chi\lm + \varphi\lm) Y\lm \right] \d r^2  + r^2 a\o^2 \left[1 + \varphi\lm  Y\lm \right] \d \Omega^2 \,, \nonumber \\ \nonumber \\ 
  \label{lltb:perturbation:2}
  \rho &= \rho^\mathrm{LTB} \left( 1 + \Delta\lm  Y\lm \right)\,, \\
  \label{lltb:perturbation:3}
  u_\mu &= \left[ u_A + \left( w\lm n_A + \frac{1}{2} k_{AB} u^B  \right) Y\lm , v\lm  Y_b\lm  \right] \,, 
 \end{align}

with sums over $(\ell, m)$ implied and $ Y_b^{(\ell m)} = \nabla_b  Y^{(\ell m)}$.\footnote{There are three types of indices appearing in the 2+2 split of the spacetime. By convention of GMG and CCF, we use Greek indices for the full spacetime coordinates, capital Roman letters for the $(t,r)$-submanifold $\mathcal{M}^2$ and small Roman letters for the angular parts on $\mathcal{S}^2$.}  The unit vectors in time and radial direction are given by $u_A = (-1,0)$ and $n_A = (0, a\p/\sqrt{1-\kappa r^2})$. $k_{AB}$ corresponds to the metric perturbation in the $(t,r)$-submanifold. \bigskip
   
Choosing the Regge-Wheeler (RW) gauge (see \cite{regge_stability_1957}), the evolution equations for the polar metric perturbations for modes $\ell \geq 2$ are then given by the closed system of master equations

\begin{align}
 \label{lltb:perturbation:4}
 \ddot{\chi} &= \frac{\chi'' - C \chi'}{Z^2} - 3 H\p \dot{\chi} + \left[A - \frac{(\ell-1) (\ell+2)}{r^2 a\o^2}\right] \chi + \frac{2\sigma}{Z} \varsigma' + \frac{2}{Z} \left[H\p ' - 2 \sigma \frac{a\p}{r a\o}\right] \varsigma - 4 \sigma \dot{\varphi} + A \varphi\,,\\
 \label{lltb:perturbation:5}
 \ddot{\varphi} &= - 4 H\o \dot{\varphi} + \left(\frac{2 \kappa}{a\o^2} - \Lambda \right) \varphi - H\o \dot{\chi} + Z^{-2} \frac{a\p}{r a\o} \chi' - \left[ \frac{1 - 2\kappa r^2}{r^2 a\o^2} + \Lambda - \frac{\ell (\ell +1)}{2r^2 a\o^2}\right] \chi + \frac{2}{Z} \frac{a\p}{r a\o} \sigma \varsigma \,, \\  
 \label{lltb:perturbation:6}
 \dot{\varsigma} &= - 2 H\p \varsigma - \frac{\chi'}{Z} \,, \\
 \label{lltb:perturbation:7}
 \eta &= 0\,.
 \end{align}

The remaining part of the field equations describes the coupling to the fluid perturbations which can be interpreted as constraints on each spatial hypersurface of constant coordinate time $t$: 

\begin{align}
  \label{lltb:perturbation:8}
 \alpha w &= \frac{1}{Z} \dot{\varphi}' - \frac{1}{Z} (\sigma - H\o) \varphi' - \frac{1}{Z} \frac{a\p}{r a\o} \dot{\chi} + \frac{H\o}{Z} \chi' + \left[ \frac{\ell(\ell+1)}{2r^2 a\o^2} + D + \frac{\kappa}{a\o^2} \right] \varsigma\,, \\
  \label{lltb:perturbation:9}
 \alpha  \Delta &= - \frac{1}{Z^2} \varphi'' + \frac{1}{Z^2} \left( C - 4 \frac{a\p}{r a\o} \right) \varphi' + \left( H\p + 2 H\o \right) \dot{\varphi} + \frac{1}{Z^2} \frac{a\p}{r a\o} \chi' + H\o \dot{\chi} \\ \nonumber
                & \quad + \left[ \frac{\ell(\ell+1)}{r^2 a\o^2} + 2 D + \Lambda \right] \left( \chi + \varphi \right) - \frac{(\ell-1)(\ell+2)}{2 r^2 a\o^2} \chi + \frac{2 H\o}{Z} \varsigma' + \frac{2}{Z} \left(H\p + H\o \right) \frac{a\p}{r a\o} \varsigma\,, \\
  \label{lltb:perturbation:10}
\alpha  v &= \dot{\varphi} + \frac{\dot{\chi}}{2} + H\p \left( \chi + \varphi \right) + \frac{1}{2 Z} \varsigma'\,.
\end{align}

The coefficients are given by the following quantities of the background $\Lambda$LTB model:

\begin{equation}
 \begin{split}
  \alpha &= 8\pi G \rho = \frac{\kappa}{a\o^2} \left( 1 + 2 \frac{a\o}{a\p}\right) - \Lambda + H\o \left( H\o + 2 H\p \right) + \frac{\kappa' r}{a\o a\p}\,, \\
  A &= 2 \alpha - \frac{6 M}{a\o^3} - 4 H\o \sigma\,, \\
  C &= \frac{a\p'}{a\p} + \frac{ \kappa r + \frac{1}{2} \kappa'r^2 }{1 - \kappa r^2} + \frac{2 a\p}{r a\o}\,, \\ 
  D &= -\frac{\alpha}{2} + H\o \left( H\o + 2H\p \right) - \Lambda\,, \\ 
  \sigma &= H\p - H\o \\
  Z &= \frac{a\p}{\sqrt{1-\kappa r^2}}\,.
 \end{split}
 \label{lltb:perturbation:11}
\end{equation}
 
Independent constraint equations can be obtained considering local energy-momentum conservation ($\nabla_\mu T^\mu_\nu=0$) which leads to dynamical equations for the fluid variables $\Delta$, $w$, and $v$ being identical to the LTB case

\begin{align}
\label{lltb:perturbation:12}
\dot{w} &= \frac{1}{2Z} \varphi' - H\p  \left( w + \frac{\varsigma}{2} \right)\,, \\
\label{lltb:perturbation:13}
\dot{\Delta} &= - \frac{\dot{\chi} + 3 \dot{\varphi}}{2} + \frac{\ell (\ell +1)}{r^2 a\o^2} v - \frac{1}{Z} \left[ \left( w + \frac{\varsigma}{2} \right)' + \left( \frac{\alpha'}{\alpha} + \frac{2 a\p}{r a\o} \right)  \left( w + \frac{\varsigma}{2} \right)\right]\,, \\
\label{lltb:perturbation:14}
\dot{v} &= \frac{\chi + \varphi}{2}\,.
\end{align} 

Regarding dipole perturbations ($\ell=1$), there is a complication as Eqs. (\ref{lltb:perturbation:4}) - (\ref{lltb:perturbation:8}) take different forms. Mathematically, there exist no dipole tensorial spherical harmonics which does not allow the trivial field equation $\eta = 0$ to hold anymore. Secondly, due to the missing tensorial components,  all perturbation variables are only partially gauge-invariant and leave an additional degree of freedom to be fixed. This issue and possible solutions are discussed in detail by GMG in \cite{gundlach_gauge-invariant_2000}. We do not want to focus on this here as we restrict our analysis to perturbations of $\ell \geq 2$. \bigskip

The axial branch is dynamically decoupled from the polar branch and is trivial for our choice of initial conditions which will be specified below. It will therefore not contribute to the numerical results presented in this work. Nonetheless we shortly describe its setup in the $\Lambda$LTB case. The linearly perturbed metric ansatz reads (see \cite{clarkson_perturbation_2009})

\begin{equation}
 \d s^2 = -\d t^2 + \frac{a\p(t,r)^2}{1-\kappa(r)r^2} \d r^2 + r^2 a\o^2(t,r) \d \Omega^2 + 2 k_A \d x^A \bar{Y}\lm_b \d x^b\,,
 \label{lltb:perturbation:15}
 \end{equation}
 
 and the axial velocity perturbation
 
 \begin{equation}
  u_\mu = \left(u_A, \bar{v} \bar{Y}\lm_a\right)\,.
  \label{lltb:perturbation:16}
 \end{equation}
 
 Defining the covariant curl $\Pi$ of the vector field $k$ given by
 
 \begin{equation}
  \Pi(t,r) = \epsilon^{AB} \nabla_B \left(\frac{k_A}{r a\o(t,r)}\right)\,, 
 \label{lltb:perturbation:17}
 \end{equation}

 the axial evolution equations for $\ell \geq 2$ in RW gauge reduce to the system
 
\begin{align}
 \label{lltb:perturbation:18}
 \dot{\bar{v}} &= 0\,, \\
 \label{lltb:perturbation:19}
 \begin{split}
 \ddot{\Pi} &= \frac{1}{Z^2} \Pi'' - \frac{\bar{C}}{Z^2} \Pi' - \left(6 H\o + H\p\right) \dot{\Pi} - \left[ 2 \alpha + 4 \Lambda + \frac{(\ell+2)(\ell-3)}{r^2 a\o^2}\right] \Pi \\
            &\ + \frac{2\alpha}{r^2 a\o^2 Z} \left( \bar{v}' + \frac{\alpha'}{\alpha} \bar{v} \right)\,, 
 \end{split} \\
  \label{lltb:perturbation:20}
  k_0 &= \frac{1}{(\ell-1)(\ell +2 )} \left[ -2 \alpha r^2 a\o^2 \bar{v} - \frac{r^4 a\o^4}{Z} \left( \Pi' + 4 \frac{a\p}{r a\o} \Pi \right) \right]\,,\\ 
 \label{lltb:perturbation:21}
  k_1 &=  \frac{1}{(\ell-1)(\ell +2 )} \left[ - r^4 a\o^4 Z \left( \dot{\Pi} + 4  H\o \Pi \right) \right]\,,  
\end{align}

 with the additional coefficient 
 
 \begin{equation}
 \bar{C} = \frac{a\p'}{a\p} + \frac{ \kappa r + \frac{1}{2} \kappa'r^2 }{1 - \kappa r^2} - \frac{6 a\p}{r a\o}\,.
  \label{lltb:perturbation:22}
 \end{equation}

\section{Initial and boundary conditions}
\label{sect:initial}

For each spherical harmonic mode $(\ell, m)$, Eqs. (\ref{lltb:perturbation:1})-(\ref{lltb:perturbation:4}) define a coupled set of linear partial differential equations in coordinate time and radius. We therefore have to specify initial and boundary conditions. The construction of boundary conditions is unaltered with respect to the previous investigations in (\cite{meyer_evolution_2015}) for the LTB case since boundary conditions are essentially defined by the geometrical properties of the solution. Since the center $r=0$ of the $\Lambda$LTB patch is an artificial boundary, certain conditions for regularity have to be applied there which have been found by GMG (see \cite{gundlach_gauge-invariant_2000})

\begin{equation*}
  \chi = \overbar{\chi} \, r^{\ell +2}\,, \ \varphi = \overbar{\varphi} \, r^{\ell}\,, \ \varsigma = \overbar{\varsigma} \, r^{\ell +1}\,, \ \Delta = \overbar{\Delta} \, r^{\ell}\,, \ w = \overbar{w} \, r^{\ell -1}\,, \ v = \overbar{v} \, r^{\ell }\,.
\end{equation*}

For $\ell \geq 2$, this fixes all perturbation variables (as well as nearly all spatial gradients) to zero at $r=0$.\footnote{ Strictly speaking, we compute the solution up to $r_\mathrm{min} \sim 1$ Mpc which is sufficiently small compared to the domain of interest of Gpc-scale} \bigskip

The outer boundary condition at $r=r_\ast$ is constructed to be causally disconnected from the domain of interest which has first been proposed in (\cite{february_evolution_2014}). By tracing null geodesics in the background $\Lambda$LTB spacetime, the exact expression for the outer boundary condition reads

\begin{equation}
  r_\ast = r_\mathrm{max} + \frac{1}{2} \int_{t_\mathrm{min}}^{t_\mathrm{max}}{\frac{\sqrt{1-\kappa(r(t))r^2(t)}}{a\p(t,r(t))} \d t}\,,
 \label{lltb:initial:1}
\end{equation}

where $r_\mathrm{max}$ is the upper bound of the domain of interest and $r(t)$ refers to the radial lightcone coordinate. According to Eq. (\ref{lltb:initial:1}), no propagating mode generated in the domain of interest and being reflected at $r_\ast$ should re-enter it within the integration time interval $\left[t_\mathrm{min}, t_\mathrm{max}\right]$. For further details and figures on the construction of boundary conditions for this setup, the reader is referred to (\cite{february_evolution_2014}) and (\cite{meyer_evolution_2015}). \bigskip   

Initial conditions are provided as radial spherical harmonic coefficient profiles on a hypersurface of constant time where the $\Lambda$LTB patch is assumed to be sufficiently close to the homogeneous and isotopic FLRW background. The coordinate time characterizing this hypersurface corresponds to the PNC time of redshift $z=100$ on the FLRW backward lightcone. For simplicity, we choose an initial scalar (Bardeen) potential $\Psi$ on this hypersurface with a 3d power spectrum given by

\begin{equation}
  \begin{split}
  \langle \Psi(\vec{k}) \Psi^\ast(\vec{k'})\rangle &= (2\pi)^3 \, P_\Psi(k) \, \delta^{(3)}_\mathrm{D}(\vec{k}-\vec{k'}) \\
                                                   &= (2\pi)^3 \, A^2_\Psi(a) \, P_\mathcal{R}(k_0) \, T^2(k) \, \delta^{(3)}_\mathrm{D}(\vec{k}-\vec{k'})\,,
  \end{split}
 \label{lltb:initial:2}
\end{equation}

where $T(k)$ denotes the matter transfer function where the fitting formula of Eisenstein \& Hu (1998) (\cite{eisenstein_baryonic_1998}) has been applied. $P_\mathcal{R}(k)$ corresponds to the power spectrum of the comoving curvature perturbation that has been evaluated at some pivot scale $k_0$ and $A_\Psi(a)$ defines an amplitude correction of the power spectrum. $\mathcal{R}$ is overall conserved in dust FLRW cosmologies and is defined as (see \cite{brechet_first-order_2009})

\begin{equation}
 \mathcal{R} = \Psi + \frac{H(a) \dot{\Psi} + H^2(a)(1-\Omega_k(a))}{4\pi G \rho(a)}\,. 
 \label{lltb:initial:3}
\end{equation}

As it is well known, cosmological inflation constrains the total power of fluctuations in $\mathcal{R}$ to 

\begin{equation}
  \mathcal{P}_\mathcal{R}(k) = \frac{k^3}{2\pi^2} P_\mathcal{R}(k) = A_s \left(\frac{k}{k\ast}\right)^{n_s} \,, 
  \label{lltb:initial:4}
\end{equation}

with the amplitude $A_s$ and spectral index $n_s$. \bigskip

At a scale $k_0 \sim 10^{-4}$ where $T(k)\sim 1$, we find $\mathcal{P}_\mathcal{R}(k_0) = 2.737 \cdot 10^{-9}$ from the Planck 2015 results (\cite{planck_collaboration_planck_2016}) which will be used throughout this work. Eq. (\ref{lltb:initial:3}) can be used to define a time-dependent amplitude correction $A_\Psi(a)$ for the conversion of the power spectrum of the primordial curvature perturbation to the Bardeen potential. Assuming a negligible time derivative $\dot{\Psi}$, we obtain

\begin{equation}
 A_\Psi(a) = \frac{1.5 \, \Omega_m(a) +  2\, \Omega_r(a)}{ 1.5\, \Omega_m(a) + 2 \, \Omega_r(a) + 1 - \Omega_k(a)}\,,
 \label{lltb:initial:5}
\end{equation}

which reduces to the well known conversion factors of $3/5$ in case of matter domination and $2/3$ in case of radiation domination. In fact, $A_\Psi$ is very close to the EdS value and radiation can still safely be described as a small test field ($\Omega_r(a=10^{-2}) \sim 2\%$) on the initial hypersurface. \bigskip

Spherical harmonic coefficient profiles $\Psi\lm(r)$ are obtained by multivariate Gaussian sampling of the spherical harmonic coefficients with a covariance matrix given by the theoretical angular power spectra. We start with a multivariate Gaussian distribution given by

\begin{equation}
 \mathcal{P}( \vec{y} ) = \frac{1}{\sqrt{ (2\pi)^n \det C }} \exp{ \left[ -\frac{1}{2} \, \vec{y}^{\,T} \cdot  C^{-1} \cdot \vec{y} \right] }
 \label{lltb:initial:6}
\end{equation}

for $n$-dimensional vectors $\vec{y} = \left\{ y_i \right\}_{0\leq i \leq n}$ and the corresponding covariance matrix

\begin{equation}
 C_{ij} = \langle y_i y_j \rangle\,.
 \label{lltb:initial:7}
\end{equation}

In order to obtain a finite realisation with the underlying distribution of Eq. (\ref{lltb:initial:3}), we first draw a vector $\vec{x}$ of $n$ uncorrelated random numbers $x_i$ with unit variance. Uncorrelated random numbers can be transformed to correlated ones by rotation in data space:

\begin{equation}
 y_i = \sum_j A_{ij} x_j\,. 
 \label{lltb:initial:8}
\end{equation}

where the coefficient matrix $A$ is determined by the Cholesky decomposition of the covariance matrix ($C=A\cdot A^T$).
 
In the particular case of spherical harmonic coefficients, the covariance matrix is given by

\begin{equation}
 C_{ij} = \left\langle \Psi\lm(r_i) \Psi\lmpast(r_j) \right\rangle\,.
\label{lltb:initial:9}
\end{equation}

Correspondingly, the vector components are a priori random numbers correlated in the radius and in all spherical harmonic modes $(\ell, m)$. \bigskip
 
In case of spatial flatness which is assumed here as first approximation for simplicity,  Eq. (\ref{lltb:initial:14}) can be expressed in terms of spherical Bessel functions $j_\ell(x)$. Using the Rayleigh decomposition of plane waves

\begin{equation}
 e^{\mathrm{i} \vec{k} \cdot \vec{r}} = 4 \pi \sum_{\ell, m} \, \mathrm{i}^\ell \, j_\ell(kf(r)) \, Y\lm(\hat{r})  \, Y\lmast(\hat{k})\,,
 \label{lltb:initial:10}
\end{equation}

with $\vec{k}= k\cdot \hat{k}$ and $\vec{r}= f(r) \cdot \hat{r}$, the spherical harmonic coefficients can be obtained as

\begin{equation}
 \Psi\lm(r) = \frac{\mathrm{i}^\ell}{2 \pi^2} \int{ \d^3k \, \Psi(\vec{k}) \, j_\ell(kf(r)) \, Y\lmast(\hat{k})}\,.
 \label{lltb:initial:11}
\end{equation}
 
A small correction $f(r)$ has to be applied here as the $\Lambda$LTB radial coordinate does not exactly match the radial coordinate of a spatially flat $\Lambda$CDM model (see \cite{february_galaxy_2013}). By comparing the coordinate-invariant proper distances in both models we obtain

\begin{equation}
 d_p^{\Lambda\mathrm{CDM}}(r) = d^{\Lambda\mathrm{LTB}}_p(r) \quad \Rightarrow  a(t_\mathrm{ini}) \, r_{\Lambda\mathrm{CDM}} = \int_0^r { \d r \frac{a\p(t_\mathrm{ini},r)}{\sqrt{1-\kappa(r)r^2}} }\,, 
 \label{lltb:initial:12}
\end{equation}

such that 

\begin{equation}
f(r) = \frac{1}{1+z_\mathrm{ini}} \int_0^r{ \d r \frac{a\p(t_\mathrm{ini},r)}{\sqrt{1-\kappa(r)r^2}} }\,.
 \label{lltb:initial:13}
\end{equation}

We can use this result to compute the covariance matrix 

\begin{equation}
\label{lltb:initial:14}
\begin{split}
\left\langle \Psi\lm(r) \Psi\lmpast(r') \right\rangle &= \frac{2}{\pi} \int_0^\infty \d k \, k^2 P_\Psi(k) \, j_\ell( kf(r) ) \, j_\ell(kf(r')) \cdot \delta_{\ell\ell'}\delta_{mm'} \\
                                                        &= C^\ell( r, r' ) \cdot \delta_{\ell\ell'} \delta_{mm'}\,.
 \end{split}
\end{equation}

 According to Eq. (\ref{lltb:initial:14}), the covariance matrix decouples into separate blocks for each spherical harmonic mode $(\ell, m)$ which only contain the radial correlations for given $\ell$-mode, i.\,e.

\begin{equation}
 C^\ell(r_i, r_j) =  \frac{2}{\pi} \int_0^\infty \d k \, k^2 \, P_\Psi(k) \, j_\ell( kf(r_i) ) \, j_\ell( kf(r_j) )\,.
\label{lltb:initial:15}
\end{equation}

The numerical approximation of integral expressions like Eq. (\ref{lltb:initial:9}) is very challenging and expensive with standard quadrature techniques. Spherical Bessel functions show a rapidly oscillatory behaviour which requires a considerable amount of function evaluations to reach acceptable accuracies. An alternative approach was proposed by Levin (1996) in \cite{levin_fast_1996, levin_analysis_1997}. In fact, the evaluation of oscillatory integrals is mapped to the problem of solving an ordinary differential equation system with no boundary conditions. The latter can be treated very efficiently by polynomial collocation. A brief sketch of this approach and its application to Eq. (\ref{lltb:initial:9}) are given in Appendix (\ref{app:levin}). \bigskip

The sampling process for each spherical harmonic mode $\ell$ can be summarized as follows:
 
 \begin{enumerate}
  \item We compute the covariance matrix $C^\ell_{ij}$ with $j \leq i$ for radial positions $r_i$, $r_j$.  
  \item The coefficient matrix $A_{ij}$ is obtained by Cholesky decomposition of the covariance matrix.
  \item We draw $2\ell+1$ uncorrelated Gaussian random numbers with unit variance for $m=0$ and variance $0.5$ for each positive orientation $m$.
  \item These uncorrelated variables can be transformed to the corresponding, radially correlated, random variables $\Psi\lm_i$ by linear combination
  \begin{equation*}
   \Psi\lm(r_i) = \sum_j A_{ij} x^{(m)}_j\,. 
  \end{equation*}
 \end{enumerate}

We restrict ourselves to orientations larger or equal zero, as these modes already contain the full information of a real-valued function on that angular scale\footnote{In fact, spherical harmonic coefficients of real-valued functions obey $a^{(\ell, -m)} = (-1)^m (a\lm)^\ast$ and therefore coefficients with negative orientations $m$ do not contain any additional degrees of freedom.}. In total, this requires us to draw $2\ell+1$ random numbers for a given $\ell$-mode. \bigskip

  \begin{figure}[!ht]
  \centering
  \subfloat[$\ell=2$]{\includegraphics[width=0.48\hsize]{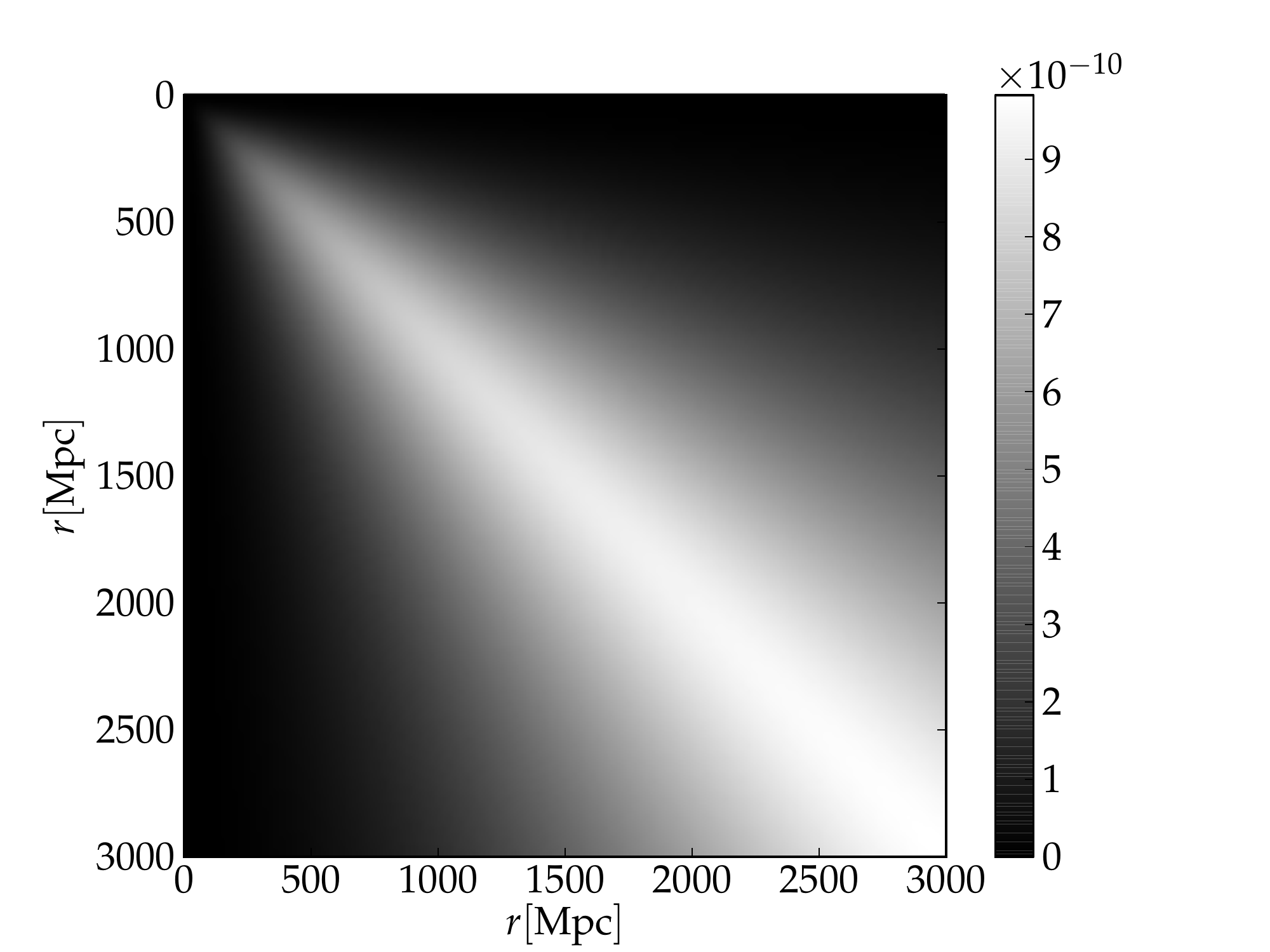} } \hfill
  \subfloat[$\ell=10$]{\includegraphics[width=0.48\hsize]{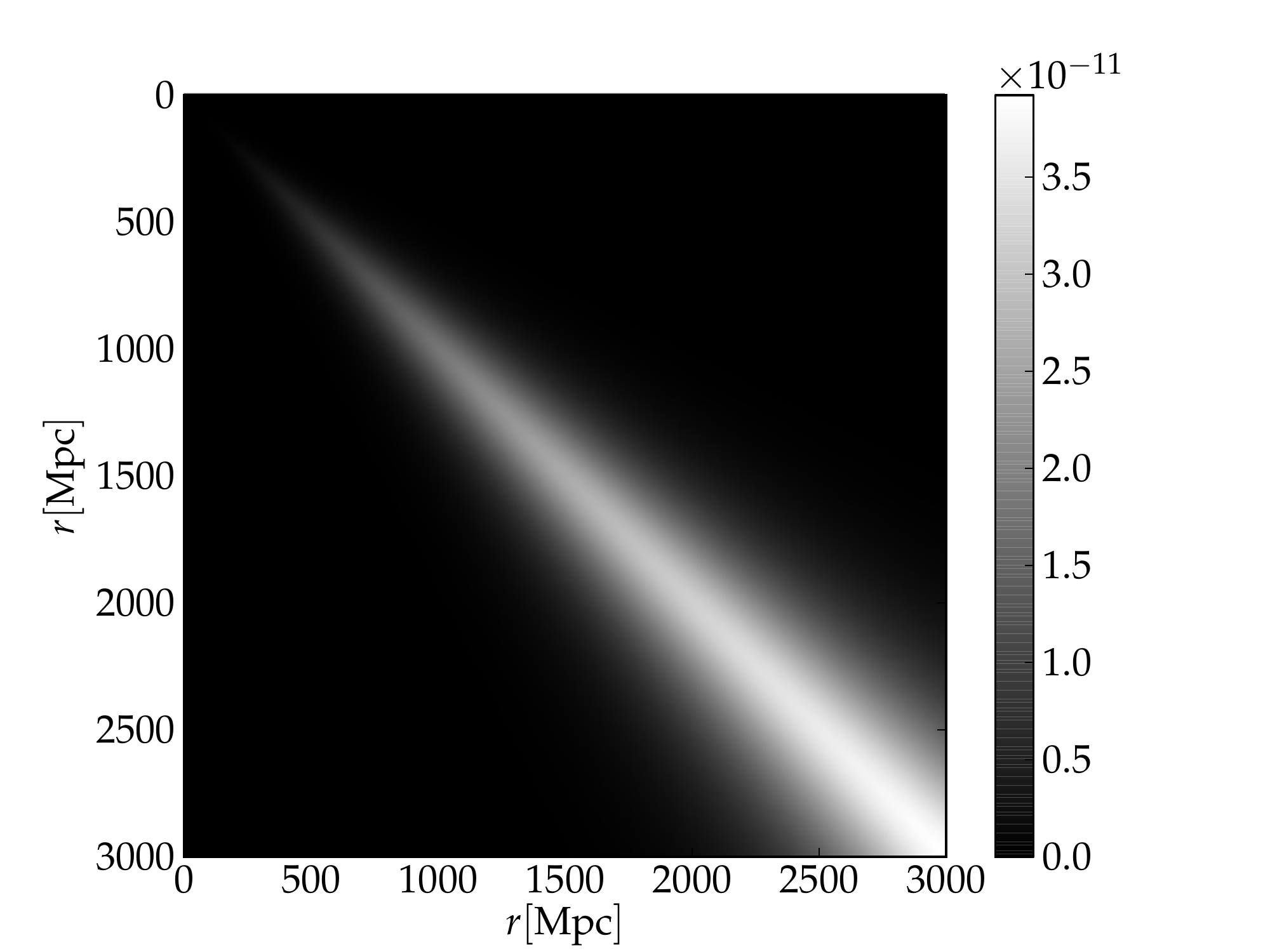} } \\
  \subfloat[$\ell=100$]{\includegraphics[width=0.48\hsize]{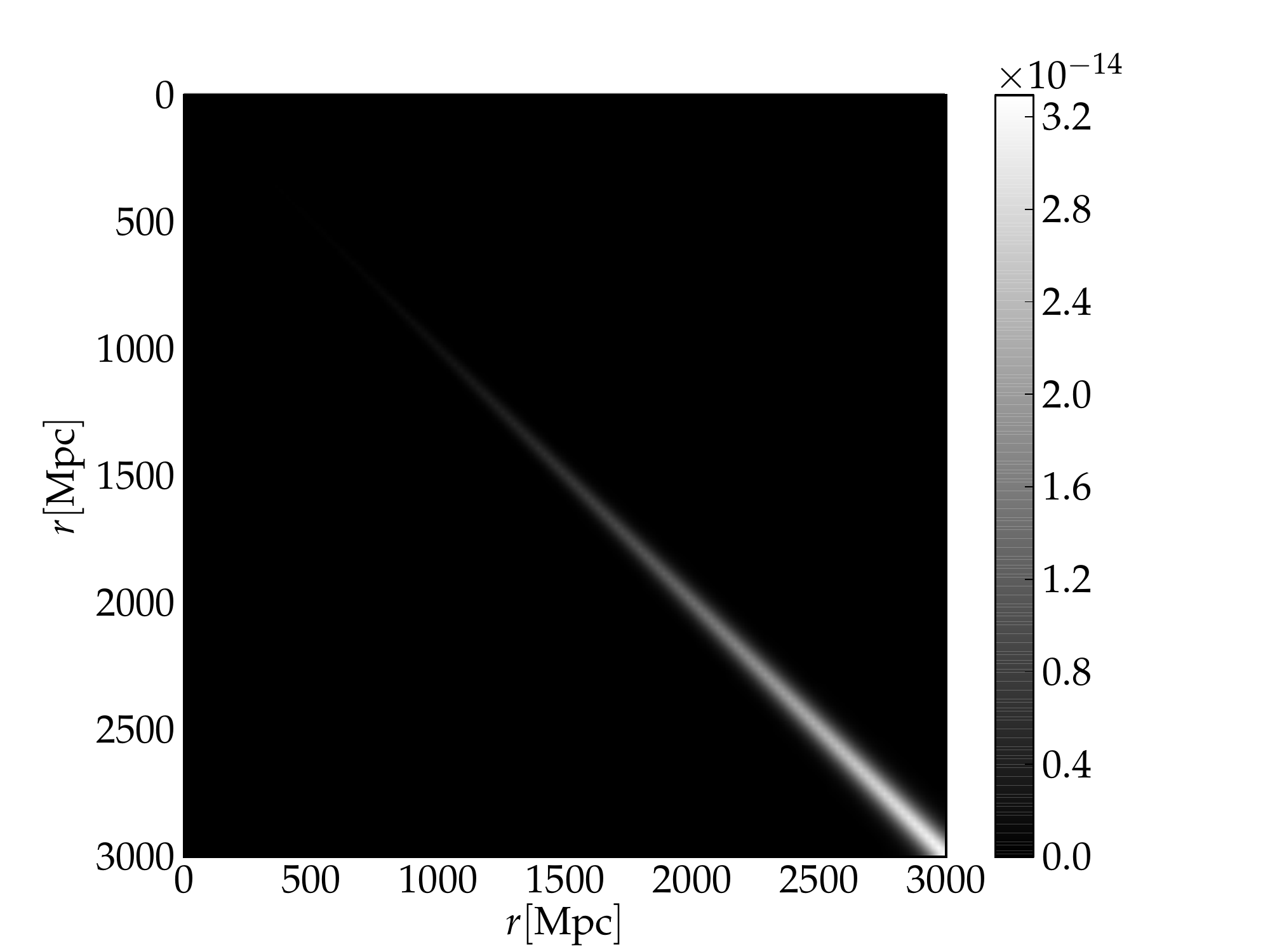} } \hfill
  \subfloat[$\ell=1000$]{\includegraphics[width=0.48\hsize]{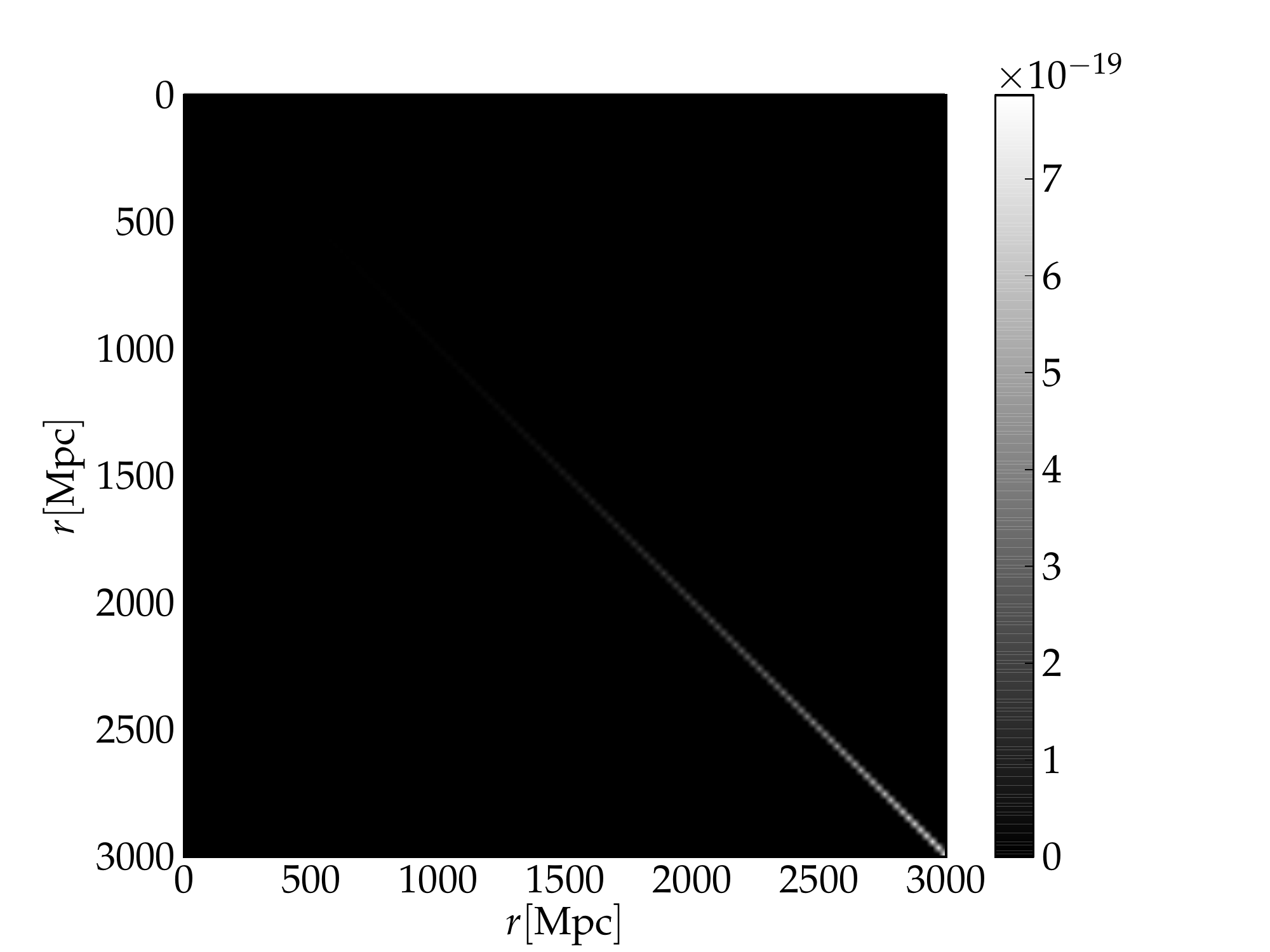} } 
  \caption[Covariance matrix entries $C^\ell_{ij}$]{These 2d plots show the covariance matrix entries that are assigned to the corresponding radii $r_i$ and $r_j$ given in Mpc. All matrices are diagonally dominant with amplitudes decreasing with $\ell$ by almost seven orders of magnitude. The latter is caused by the strong decay of the initial power spectrum at large $k$-modes ($P_\Psi(k) \sim k^{-7}$). The radial correlation increases with distance from the center which is expected as structures of given angular scale must have larger tangential extensions. This itself leads to larger radial scales in a statistically isotropic initial universe. Since small $\ell$-modes describe angular patches corresponding to larger fluctuations in spatial scale, the radial correlation also increases for those modes correspondingly.}
 \label{lltb:initial:fig:1}
  \end{figure}
 
   \begin{figure}[!ht]
  \centering
    \subfloat[$r=r_\mathrm{max}/2 - \Delta r$]{\includegraphics[width=0.6\hsize]{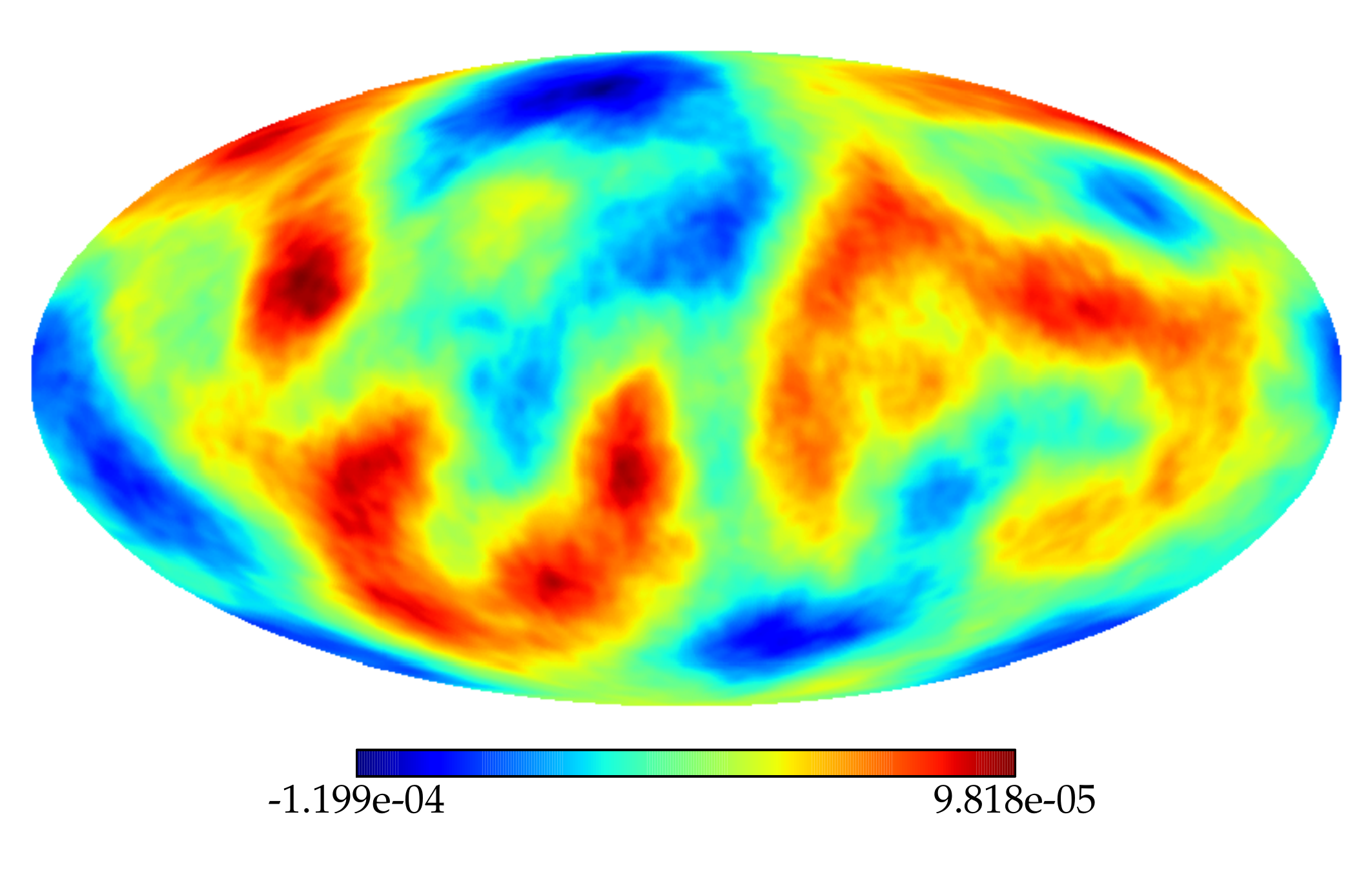} }  \\
    \subfloat[$r=r_\mathrm{max}/2$]{\includegraphics[width=0.6\hsize]{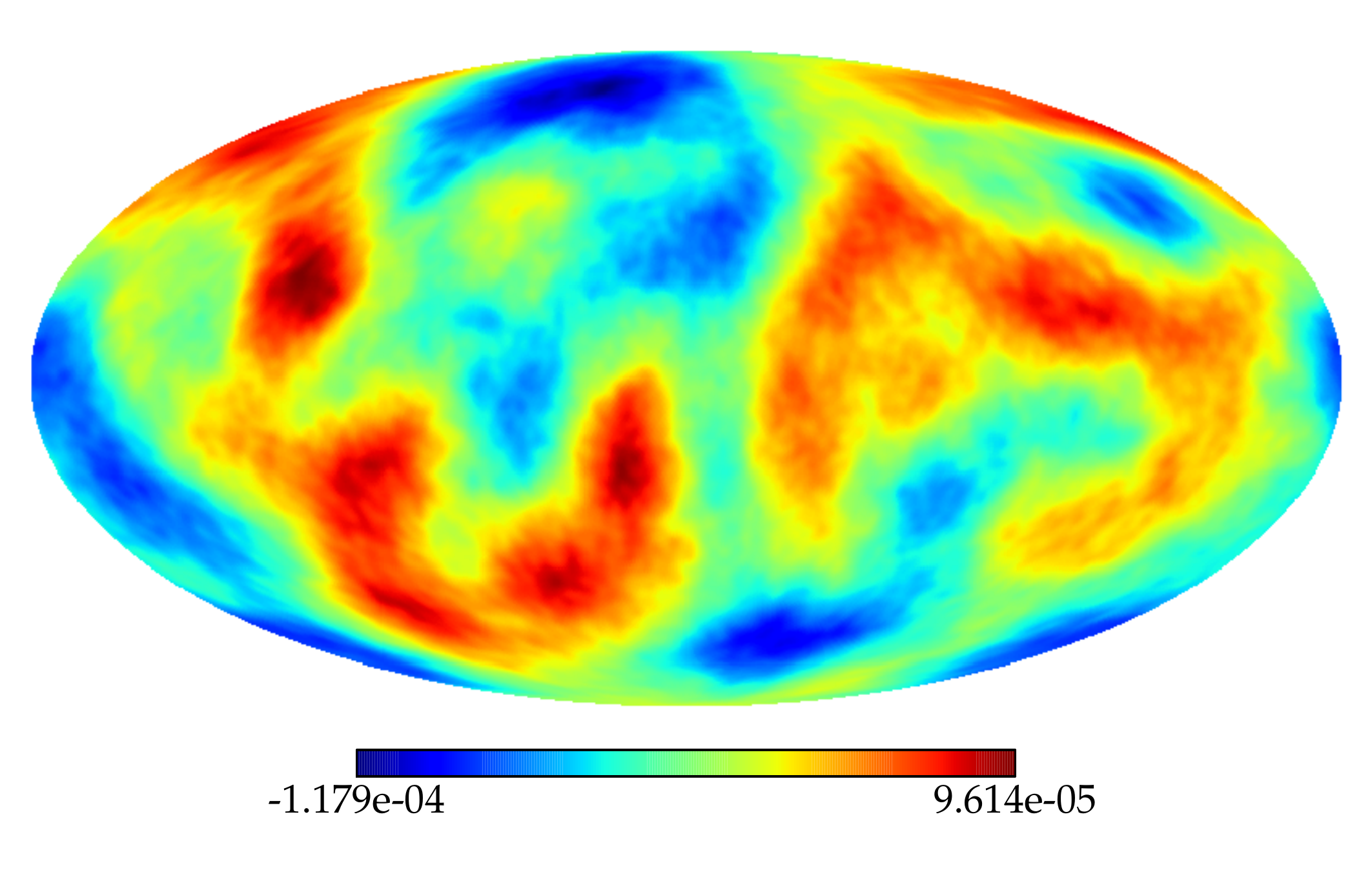} } \\
    \subfloat[$r=r_\mathrm{max}/2 + \Delta r$]{\includegraphics[width=0.6\hsize]{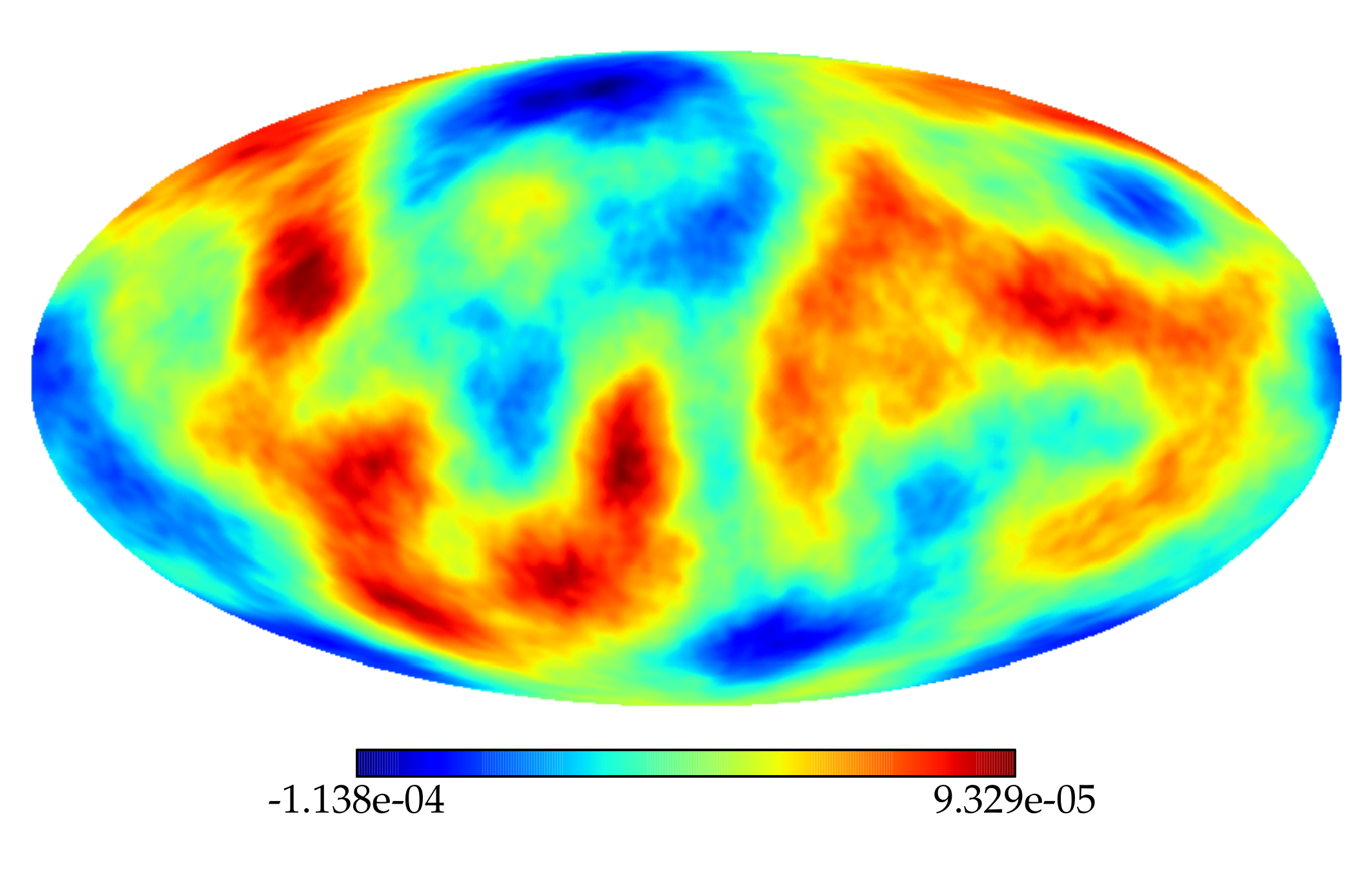} } 
  \caption[Consecutive Healpix Maps around a fiducial radius]{These figures show consecutive Healpix Maps around a fiducial radius $r=r_\mathrm{max}/2 = 1500 \ \mathrm{Mpc}$ with $\Delta r \sim 47 \ \mathrm{Mpc}$. The correlation of fluctuations on large angular scale can clearly be seen.}
  \label{lltb:initial:fig:2}
  \end{figure}

The covariance matrices for different $\ell$-modes are shown in Fig. (\ref{lltb:initial:fig:1}). The correlation is expected to increase with radius as fluctuations at large distances from the center have to be larger in spatial scale in order to appear under the same angle. Fig. (\ref{lltb:initial:fig:2}) shows Healpix maps of spherical shells at close radial bins. Considering maps of increasing radius, one can see that potential fluctuations are indeed correlated. This correlation is more prominent on large angular scales as indicated by the shape of the covariance matrices. \bigskip
  
 As shown by Clarkson et al. (2009) in \cite{clarkson_perturbation_2009}, the only remaining gauge-invariant polar perturbation in the FLRW limit with initial scalar perturbations is $\varphi\lm = -2 \Psi\lm$. This result is based on the geometrical construction of gauge-invariant perturbations in spherically symmetric dust spacetimes and therefore also holds in $\Lambda$LTB models. We therefore obtain the simple initial configuration
 
 \begin{equation}
  \begin{split}
   &\varphi\lm(t_\mathrm{ini},r) = -2 \Psi\lm(t_\mathrm{ini},r)\,, \\
   &\chi\lm(t_\mathrm{ini},r) = 0 = \varsigma\lm(t_\mathrm{ini},r)\,, \\
   &\dot{\chi}\lm(t_\mathrm{ini},r) = 0 = \dot{\varphi}\lm(t_\mathrm{ini},r)\,, \ \dot{\varsigma}\lm(t_\mathrm{ini},r) = 0\,,
  \end{split}
  \label{lltb:initial:17}
\end{equation}

for the metric perturbations which respect the grid structure and boundary conditions posed by the problem itself. The initial fluid perturbations ($\Delta\lm, w\lm, v\lm $) are then constrained by Eqs. (\ref{lltb:perturbation:5}) - (\ref{lltb:perturbation:7}). \bigskip
  
All perturbation variables have to vanish initially in the region $[r_\mathrm{max}, r\st ]$. We define a transition region $[r_\mathrm{max}, r_\mathrm{ext}]$ where the initial profile is extended by a Gaussian function centered at $r_\mathrm{max}$ and a FWHM of one fifth of the size of the extension region. For details we refer to (\cite{meyer_evolution_2015}).

\section{Numerical setup}
\label{sect:numeric}

The numerical integration of the polar master equation system (\ref{lltb:perturbation:1})-(\ref{lltb:perturbation:4}) and consecutive evaluation of the constraint equations (\ref{lltb:perturbation:5})-(\ref{lltb:perturbation:8}) is done with the help of the Distributed Unified Numerics Environment (DUNE) (see \cite{bastian_towards_2004, bastian_generic_2008-1, blatt_generic_2008, bastian_dune_2011}) which has already been applied and discussed in detail in (\cite{meyer_evolution_2015}). We therefore just give a short summary here. \bigskip

We employ the method of lines leaving the time coordinate continuous and discretizing the radial coordinate using finite elements. The latter turns out to be more flexible and stable than finite differences and does not suffer from instabilities close to the center of the $\Lambda$LTB patch. The resulting coupled large scale ODE problem is then integrated with a third order Alexander S-stable diagonally implicit time integration scheme (see \cite{alexander_diagonally_1977}) which has necessary stability properties also on small angular scales $\ell$. An implicit time integration scheme is, of course, less efficient than an explicit one that would typically be applied. In fact, it turns out that a finite difference implementation with explicit time integration is very efficient at small $\ell$ modes (see \cite{february_evolution_2014, gundlach_critical_2002}), but we think that  on small angular scales, the contributions to Eq. (\ref{lltb:perturbation:1}) proportional to $\ell^2/r^2$ cause severe numerical stiffness of Eq. (\ref{lltb:perturbation:1}) which requires an implicit solver in order to avoid strong restrictions on the size of  timesteps. We also applied partially implicit solvers like the recently developed PIRK methods (see \cite{cordero-carrion_partially_2012}) which turned out to be more robust, but could not alleviate those restrictions in a sufficient manner. \bigskip

Although an implicit solver is not limited to the Courant-Friedrics-Levy condition (\cite{courant_uber_1928}), we nonetheless adapt the timesteps according to that condition, because characteristics of the system define its natural timescale. Hence, we choose

\begin{equation}
\frac{\Delta t(t)}{\Delta r} = \min_{ r_\mathrm{min}\le r \le r_\mathrm{max}} (Z(t,r)) 
\label{lltb:numerics:1}
\end{equation}

The background model coefficients given in Eq. (\ref{lltb:perturbation:11}) are precomputed and evaluated exactly at the grid points and timesteps given by Eq. (\ref{lltb:numerics:1}). \bigskip 

Given a set of angular scales $\{\ell\}$, Eqs. (\ref{lltb:perturbation:1})-(\ref{lltb:perturbation:8}) are evolved for all $\ell+1$ possible orientations $m \geq 0$. Each timestep defines a spatial hypersurface that intersects the $\Lambda$LTB backward lightcone. The resulting spherical harmonic coefficient set is evaluated at these intersections stored as function of the corresponding redshift bin on the central observer's past null cone defined by Eqs. (\ref{lltb:background:10}) and (\ref{lltb:background:11}).

\section{Angular Power spectra and Coupling strength}
\label{sect:coupling strength}

Spherical harmonic power spectra of each metric and fluid variable and the corresponding cosmic variance limit can be estimated by
\begin{align}
 \label{lltb:coupling strength:1}
  C^\ell_X(z) &= \frac{1}{2\ell+1} \sum_{m=-\ell}^{\ell} \left| a_X\lm(t(z), r(z))\right|^2\,, \\
 \label{lltb:coupling strength:2}
  \Delta C^\ell_X(z) &= \frac{2\ell+1}{2} C^\ell_X(z)\,,
\end{align}

where $X \in \left\{ \chi, \varphi, \varsigma, \Delta, w, v \right\}$. \bigskip

Due to dynamical coupling of the gauge-invariant metric variables we expect an initial FLRW scalar $\sim \varphi\lm$ to create non-vanishing and possibly significant contributions of the two initially vanishing gauge-invariant variables $\chi\lm$ and $\varsigma\lm$. Those influence the evolution of $\varphi\lm$ and the fluid variables $\Delta\lm$, $w\lm$, and $v\lm$. We therefore estimate the coupling strength $\epsilon^\ell$ by comparing the estimated angular power spectra of $\varphi\lm$ and $\Delta\lm$ of the coupled and uncoupled evolution and express their absolute deviation in units of the cosmic variance of the uncoupled evolution. Thus, we obtain

\begin{equation}
 \epsilon^\ell_X(z) = \frac{2}{2\ell+1}\frac{ \left|C^\ell_X(z) - C^\ell_{X,\mathrm{uc}}(z)\right|}{ C^\ell_{X,\mathrm{uc}}(z)}\,.
 \label{lltb:coupling strength:3}
\end{equation}

The definition of Eq. (\ref{lltb:coupling strength:3}) for $X \in \left\{\varphi, \Delta \right\}$ allows to conclude whether the influence of coupling is significant with respect to the cosmic variance limit. \bigskip

We finally average over all $\ell$ modes considered in order to estimate a mean coupling strength at each redshift bin for the given set of $\ell$-modes. Although its absolute value depends on that particular set of angular scales $\ell$ considered in the analysis, it helps to show the dependence of the coupling strength as a function of redshift and therefore PNC position in the $\Lambda$LTB patch.

\section{Results and Discussion}
\label{sect:results}

The ideas outlined in the previous sections can now readily be applied to an arbitrary $\Lambda$LTB cosmology for which the angular power spectra and coupling strengths can be extracted on the central past null cone. We define generic $(\Lambda)$LTB model by six parameters that fix the asymptotic FLRW model as well as a set of three equidistant nodes $\{a_i\}$ for its radial profile in the domain of interest. Since the $\Lambda$LTB model matches its asymptotic FLRW model at large redshifts, this also fixes the initial set of gauge-invariant perturbations. In a cosmologically relevant case, we are particularly interested in results for background models that have already been constrained by observational data that do not assume any information from linear structure formation. Those have reliably been estimated in Redlich et al. (2014) (\cite{redlich_probing_2014}). We will test the evolution of gauge-invariant linear perturbations for two spatially inhomogenous models and a reference homogeneous FLRW model: 

\begin{itemize}
 \item the \textit{best fit $\Lambda$LTB model} (bf$\Lambda$LTB) constrained by measurements of the local Hubble rate, distance redshift relations of type IA supernovae, the CMB spectrum and upper bounds of the kinetic Sunyaev-Zel'dovich effect
 \item the \textit{best fit LTB model} (bfLTB) constrained by distance redshift relations given the local Hubble rate and type IA supernovae. We consider this model for completeness though it does not fit the CMB spectrum appropriately
 \item a \textit{reference $\Lambda$CDM model} (ref$\Lambda$CDM) sharing the same background cosmological parameters with the bf$\Lambda$LTB model
\end{itemize}

\begin{table}
\centering
\begin{tabular}{l|rrrrrr}
  \toprule \\
  model & $h$ & $\Omega_\mathrm{m}$ & $\Omega_\Lambda$ & $a_1$ & $a_2$ & $a_3$ \\ 
  \midrule \\
  bf$\Lambda$LTB &  0.73 & 0.245 & 0.745 &  1.02 & 1.02 & 0.96 \\
  bfLTB &  0.557 & 1.0 & 0.0 &  0.23 & 0.44 & 0.59  \\
  ref$\Lambda$CDM & 0.73 & 0.245 & 0.745 & 1.0 & 1.0 & 1.0 \\
 \bottomrule
 \end{tabular}
 \caption{Table containing the model parameters of the background models considered in this work.}
 \label{lltb:results:tab:1}
\end{table}

\begin{figure}[!ht]
  \subfloat[bf$\Lambda$LTB, $z=0.1$ ]{ \includegraphics[page=1, width=0.47\hsize]{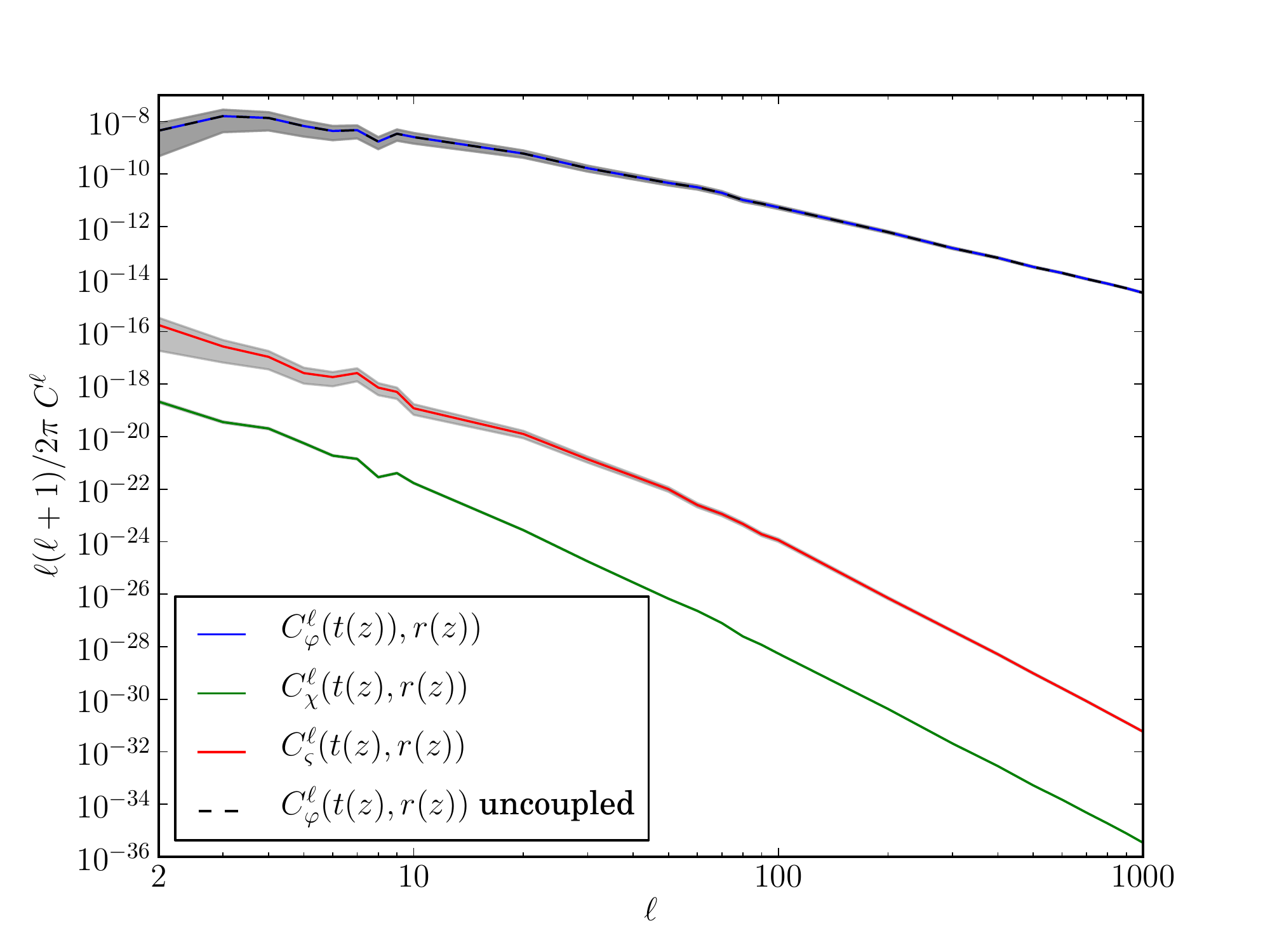} } \hfill
  \subfloat[bfLTB, $z=0.1$]{ \includegraphics[page=1, width=0.47\hsize]{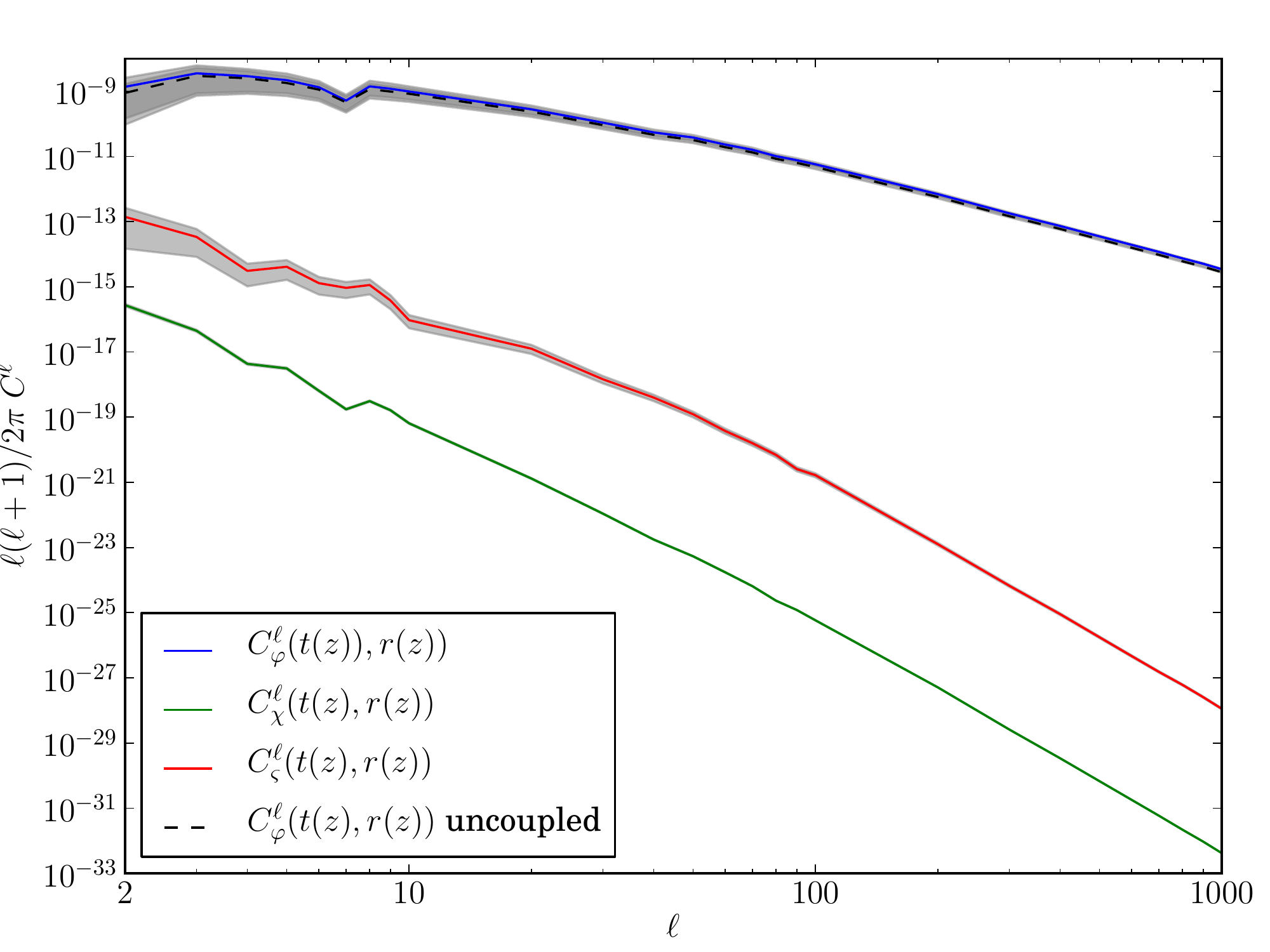} } \\
  \subfloat[bf$\Lambda$LTB, $z=0.3$ ]{ \includegraphics[page=3, width=0.47\hsize]{powerspectra_LambdaLTB.pdf} } \hfill
  \subfloat[bfLTB, $z=0.3$]{ \includegraphics[page=3, width=0.47\hsize]{powerspectra_LTB.pdf} } \\
  \subfloat[bf$\Lambda$LTB, $z=0.5$ ]{ \includegraphics[page=5, width=0.47\hsize]{powerspectra_LambdaLTB.pdf} } \hfill
  \subfloat[bfLTB, $z=0.5$]{ \includegraphics[page=5, width=0.47\hsize]{powerspectra_LTB.pdf} } \\
  \caption{Angular power spectra of the three metric variables in the bf$\Lambda$LTB and bfLTB models according to Tab (\ref{lltb:results:tab:1}) for a sample of three redshift bins. In both cases the initially zero variables are dynamically generated during spacetime evolution. The spectrum of the freely evolved metric variable $\varphi\lm$ is plotted as well to illustrate the influence of coupling effects on the dynamical behaviour. In the bf$\Lambda$LTB case, there is no noticeable effect of coupling and the coupling terms lies within the cosmic variance limit whereas the void model shows substantial deviations of the fully coupled and freely evolved solution as expected.}
  \label{lltb:results:fig:1}
 \end{figure}

The model parameters $\{h, \Omega_\mathrm{m}, \Omega_\Lambda, a_1, a_2, a_3 \}$ are shown in Tab. (\ref{lltb:results:tab:1}). All models are studied in a domain of interest around the center having a radial extent of $3$ Gpc. The $\Lambda$LTB model is very close to a $\Lambda$CDM model with only percent level deviations of the background density from a spatially homogeneous form whereas the best fit LTB model has the expected underdense shape that leads to an increase of the local Hubble rate allowing to match distance redshift relations of type Ia supernovae. The angular power spectra of all metric variables $\chi$, $\varphi$ and $\varsigma$ of the two best fit spatially inhomogeneous models are shown in Fig. (\ref{lltb:results:fig:1}) for three exemplary redshift bins. In addition, we plot the solution for the initially non-vanishing metric variable $\varphi$ that has been evolved freely neglecting the influence of dynamical coupling. The redshift bins are chosen such that intersections of the past null cone with the radially inhomogeneous hypersurfaces of constant time lead to radial coordinates within the predefined domain of interest. Therefore we only consider redshifts $z \lesssim 1$ that ensure the lightcone being contained within the domain of interest. \bigskip   

We see that in both cases non-vanishing contributions of the variables $\chi$ and $\varsigma$ are generated dynamically. However, in case of the bf$\Lambda$LTB model, we see no noticeable influence of coupling within the cosmic variance limit on the evolution of $\varphi$. This is an expected due to the small deviations from a flat density profile. On the contrary, the bfLTB model shows strong influence of coupling effects that have already been found in (\cite{meyer_evolution_2015}) for a Gaussian shaped toy model for the density profile having similar size and depth. \bigskip

  \begin{figure}[!ht]
  \subfloat[$z=0.1$]{ \includegraphics[page=1, width=0.48\hsize]{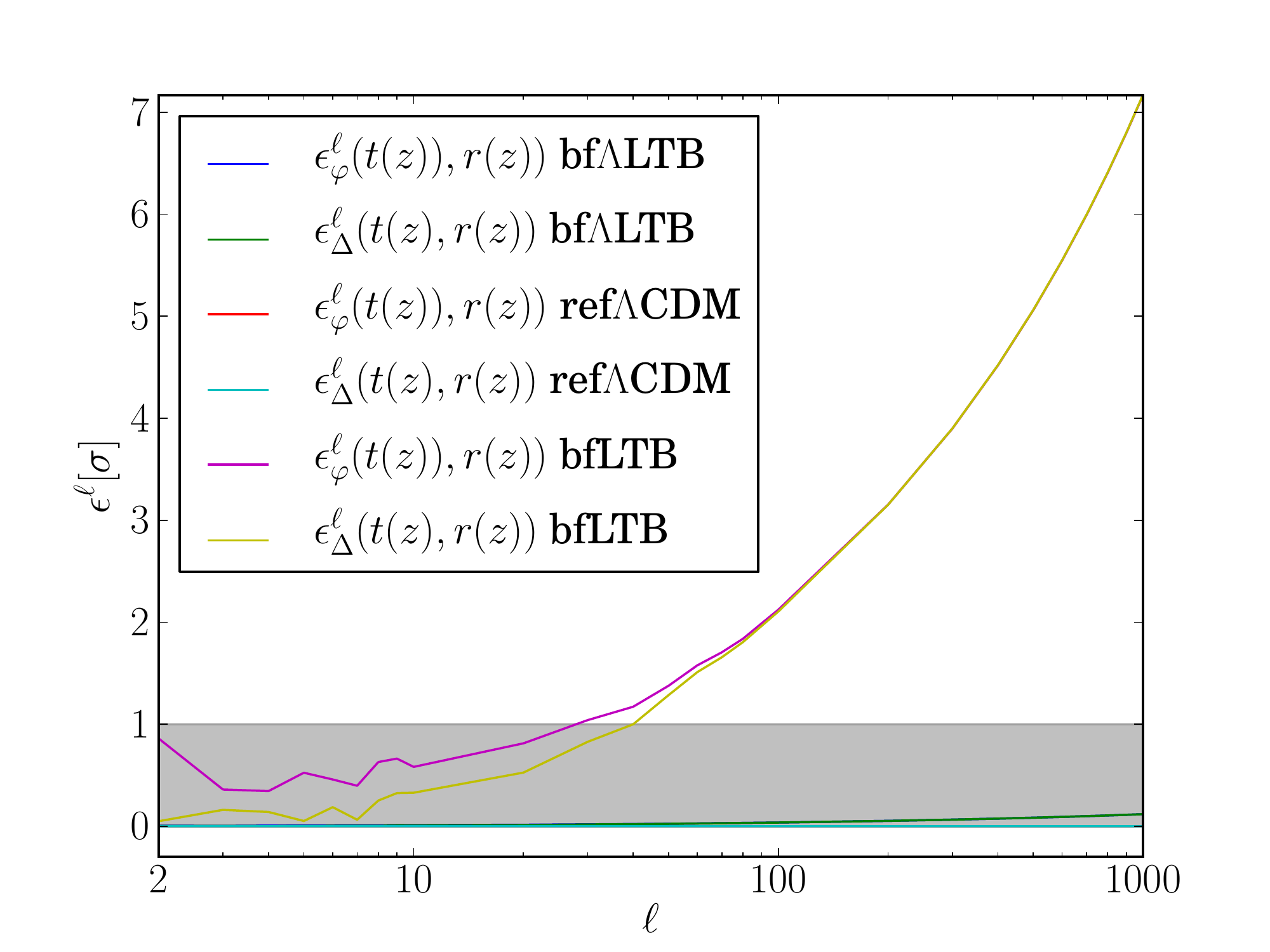} } \hfill
  \subfloat[$z=0.3$]{ \includegraphics[page=3, width=0.48\hsize]{combined_coupling_strength.pdf} } \\
  \subfloat[$z=0.5$]{ \includegraphics[page=5, width=0.48\hsize]{combined_coupling_strength.pdf} } \hfill
  \subfloat[$z=0.7$]{ \includegraphics[page=7, width=0.48\hsize]{combined_coupling_strength.pdf} } \\
  \caption{Coupling strength $\epsilon^\ell_X$ as obtained in all three models defined by Eq. (\ref{lltb:coupling strength:3}) for the variables $\varphi\lm$ and $\Delta\lm$. Results are shown as function of the angular scale for four exemplary positions on the model's central past null cone. The coupling strength is constructed in units of the expected cosmic variance limit $\sigma$ which is shown as grey shaded area. Whereas the ref$\Lambda$CDM model shows no dynamical coupling at all, the bf$\Lambda$LTB model shows small effects at large $\ell$ modes whereas coupling effects in the LTB void model are prominent features and have significant influence even on larger angular scales.  On small angular scales, the coupling strength is expected to increase quadratically in $\ell$ as coupling coefficients show the same proportionality. The curves of $\varphi$ and $\Delta$ line up for large $\ell$ as the influence of terms  $\sim \ell^2$ are dominating the coupling behaviour. Coupling effects also strongly depend on the position in the ($\Lambda$)LTB patch which is illustrated in Fig. (\ref{lltb:results:fig:3}).}
  \label{lltb:results:fig:2}
 \end{figure}

The estimated coupling strength according to Eq. (\ref{lltb:coupling strength:3}) is presented in Fig. (\ref{lltb:results:fig:2}) for all three models considered and four exemplary redshift bins that cover the domain of interest on the past null cone. We see that, as expected, the ref$\Lambda$CDM model shows no coupling at all whereas coupling effects for the bf$\Lambda$LTB model for $\varphi$ and $\Delta$ are below the cosmic variance limit and will therefore not be noticed\footnote{We have to admit that for very large $\ell$-modes at intermediate redshifts we see that the relative deviation is in fact larger than the cosmic variance limit. However, observables at large $\ell$ modes will also be affected by non-linear effects of structure formation which are not understood yet in $\Lambda$LTB models. So this region has to be treated carefully anyway in a future analysis.}.  \bigskip

As expected, coupling increases to multiples of the cosmic variance limit in the bfLTB model which confirms its strong influence on the evolution on the metric and fluid variables. In each case, we see a quadratic increase of coupling with $\ell$. Coefficients $\sim \ell(\ell+1)$ in Eqs. (\ref{lltb:perturbation:1})-(\ref{lltb:perturbation:8}) dominate the coupling terms in that regime. For the same reason, the results $\epsilon^\ell_\varphi$ and $\epsilon^\ell_\Delta$ line up as the influence of the initially vanishing metric variables $\chi$ and $\varsigma$  is subdominant with respect to $\varphi$. Therefore, we see that $\Delta$ in Eq. (\ref{lltb:perturbation:5}) is mainly sourced by a term $\ell^2 \varphi\lm$. This way, the coupling effects on $\varphi$ are directly mapped to $\Delta\lm$ and the curves line up for large $\ell$-modes.\bigskip  

 \begin{figure}[!ht]
  \centering
  \subfloat[bf$\Lambda$LTB]{ \includegraphics[width=0.48\hsize]{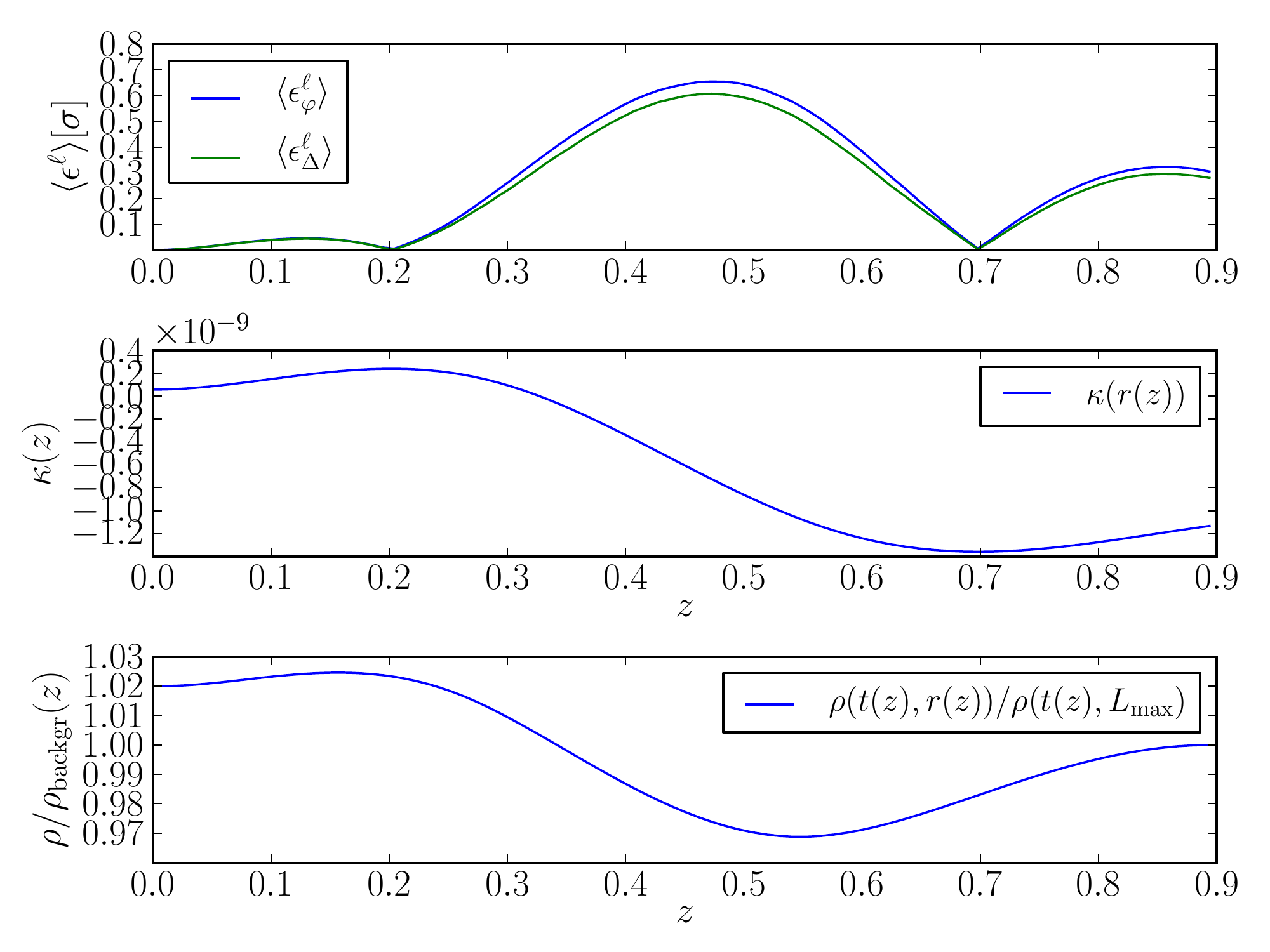} } \hfill
  \subfloat[bfLTB]{ \includegraphics[width=0.48\hsize]{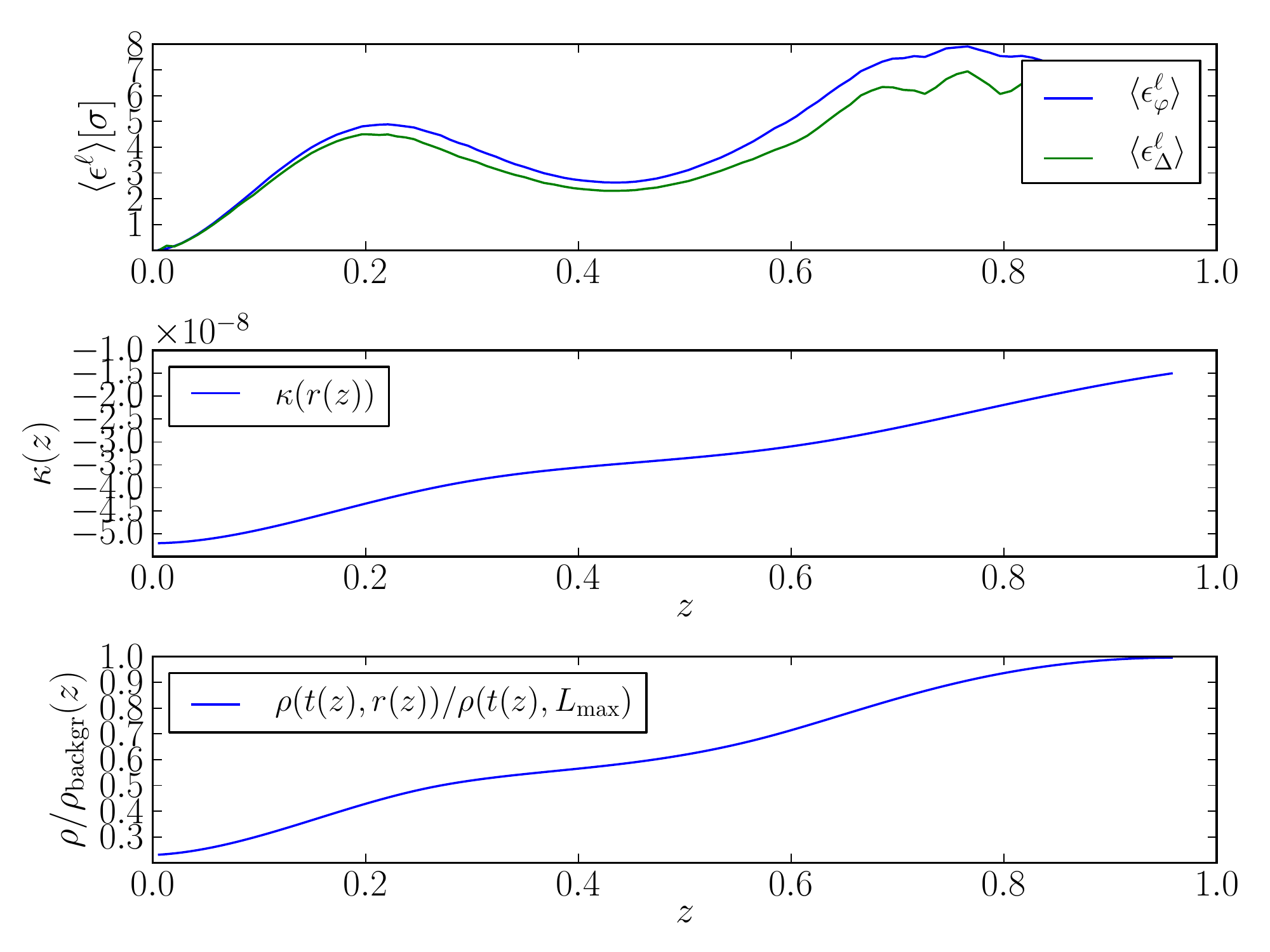} } \\
  \subfloat[ref$\Lambda$CDM]{ \includegraphics[width=0.48\hsize]{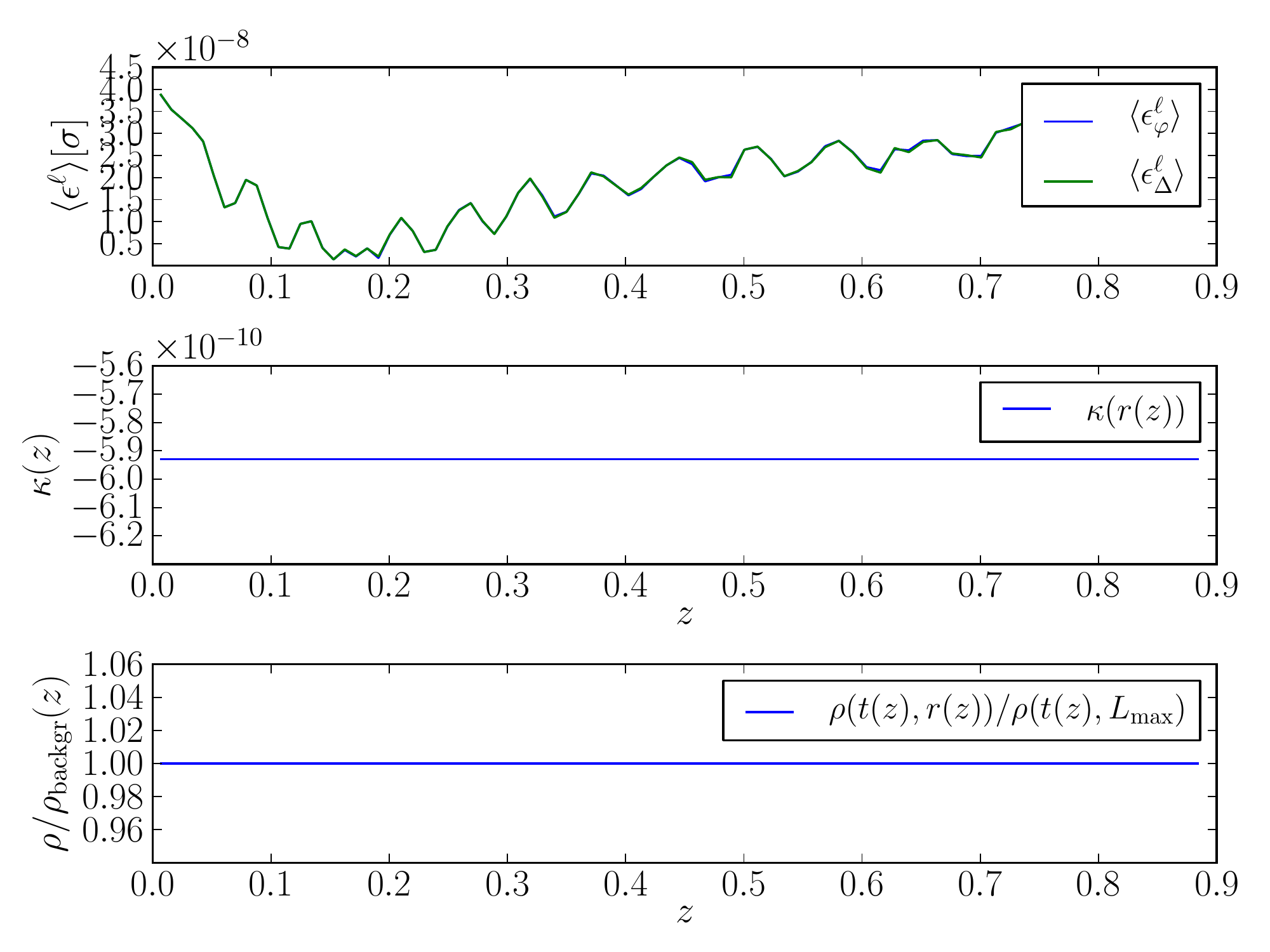} } \\
  \caption{$\ell$-averaged coupling strength $\langle\epsilon^\ell_X\rangle_\ell$ for $\varphi$ and $\Delta$ as function of redshift in combination with the curvature- and density profile of the inhomogeneous patch evaluated on the past null cone. It can be seen that positions with strong gradients in the curvature and density profile also show prominent coupling effects. This is caused by strong background shear effects sourced by spatial gradients in the curvature profile $\kappa(r)$. Although the absolute value of the averaged coupling strength dependens on the set of $\ell$-modes considered in the analysis, the dominant coupling effects can clearly be seen in case of the bfLTB void model whereas, within the cosmic variance limit, it has a non-noticeable effect in case of $\Lambda$LTB models. The spatially homogeneous ref$\Lambda$CDM model has zero dynamical coupling of the gauge-invariant $\varphi$ to  $\chi$ and $\varsigma$ and therefore serves as crosscheck for any artificial coupling generated by the numerical evolution. This estimated ``artificial" coupling strength is of the order of $10^{-8}$ of the cosmic variance limit.}
  \label{lltb:results:fig:3}
 \end{figure}
 
The redshift dependence of the coupling effect can be seen in Fig. (\ref{lltb:results:fig:3}). The $\ell$-averaged coupling strength is shown as function of redshift in the ($\Lambda$)LTB patch. As already mentioned in Sect. (\ref{sect:coupling strength}), the absolute value of the $\ell$-averaged coupling strongly depends on the set of modes considered in the investigation. Nonetheless, its relative change with redshift illustrates the dependence of coupling effects at different positions in the radially inhomogeneous patch. In comparison to the curvature and density profiles we see that the coupling strength is enhanced at positions of strong spatial gradients in both profiles. There the spatial anisotropy ($\sigma = H\p-H\o$) in the expansion rate is maximised since it is sourced by gradients in the spatial curvature profile $\kappa(r)$ (see Eqs. (\ref{lltb:background:9}) and (\ref{lltb:background:10})). Those spatial gradients then transform into gradients in redshift of the backward lightcone. From a theoretical point of view, the reference spatially homogeneous $\Lambda$CDM model has zero dynamical coupling of the gauge-invariant $\varphi$ to  $\chi$ and $\varsigma$. It therefore serves as a crosscheck for any artificial coupling generated by the numerical evolution. The estimated ``artificial" coupling strength is of the order of $10^{-8}$ of the cosmic variance limit far below any noticeable influence on the results obtained in Figs. (\ref{lltb:results:fig:1}) and (\ref{lltb:results:fig:2}). \bigskip

Based on this investigation, we can conclude that the bf$\Lambda$LTB model has indeed negligible coupling effects on the evolution of linear gauge-invariant perturbations and dynamical coupling can therefore safely be neglected in this case. However, spherical void models that allow to fit the local distance redshift relation, show prominent coupling effects and have a strong influence on the spacetime evolution and on observables predicted from that. Fortunately, those void models have already been excluded observationally by classes of observables that do not particularly rely on linear structure formation. A radially inhomogeneous dust solution including a cosmological constant that describes our universe reliably can therefore safely be investigated by neglecting dynamical coupling effects. This leads to a substantial simplification of the problem as the free evolution of $\varphi$ is just constrained by an ordinary differential equation being independent of the spherical harmonic modes $(\ell,m)$. This allows to construct transfer functions and corresponding theoretical angular power spectra that do not suffer from cosmic variance limits.     

\FloatBarrier
  
\section{Conclusion and Outlook}
\label{sect:conclusion}

We have investigated the effects of dynamical coupling of the gauge-invariant metric and fluid variables in a radially inhomogeneous dust solution of Einstein's field equations. For this purpose, we have applied a previously developed numerical scheme to evolve the coupled master equation system that determines the evolution of these variables and pose initial spherical harmonic coefficient profiles based on multivariate Gaussian sampling of exact angular covariance matrices. We found that for cosmologically relevant choices of the radial density profile of the inhomogeneous patch, those coupling effects are negligibly small compared to the cosmic variance limit. Dynamical coupling can therefore safely be neglected in this case. On the contrary, in the bfLTB model, coupling effects are strong and have a noticeable influence on the spacetime evolution of an initially non-vanishing scalar gravitational potential. In those models, we would have to take care of these effects, but careful investigations and multi-probe analyses have shown that these models are in severe tension with observational data and are therefore excluded from a scientifically relevant description of the local universe. \bigskip 

Since best fit $\Lambda$LTB models do not need to be asymptotically flat, we plan to extend the approach in Sect. (\ref{sect:initial}) to hyperspherical Bessel functions to allow for a more reliable description of spherical harmonic profiles on the initial hypersurface. However, we expect the effect to be subdominant as the curvature radius of the asymptotic FLRW universe is still much larger than the domain of interest considered. Nonetheless, it is important to investigate its influence.  \bigskip

The final goal of this line of investigations are additional constraints on spatial inhomogeneity of $\Lambda$LTB models in the best possible cases. If we include results from linear perturbation theory, we need to construct a physically meaningful set of observables. Although, the gauge-invariant variables proposed in Clarkson et al. (2009) (see \cite{clarkson_perturbation_2009}) are physical variables and therefore possibly observable, their physical interpretation is nontrivial. We therefore plan to construct physical observables from a theory of light propagation in perturbed $\Lambda$LTB models yielding angular power spectra of well known quantities such as shear and convergence. Those investigations are currently underway and will be considered in a forthcoming paper. 

\appendix

\section{Appendix: Levin collocation method}
\label{app:levin}

Since the Levin collocation method developed in (\cite{levin_fast_1996}) is not well known in the context of cosmology and spherical harmonic analysis and has, to our knowledge, only previously been applied in (\cite{zieser_cross-correlation_2016}), we briefly sketch this approach in the following.\bigskip 

The formalism addresses integrals of the form

\begin{equation}
 I = \int_a^b{ \vec{f}(x)^T \vec{w}(x) \, \d x} = \int_a^b{ \langle \vec{f},\vec{w} \rangle \, \d x}\,,
 \label{lltb:levin:1}
\end{equation}

with $\vec{f} = \left(f_1(x), \ldots, f_m(x) \right)^T \in \mathbb{R}^m$ being a vector of $m$ non-oscillating functions and $\vec{w}(x) = \left(w_1(x), \ldots, w_m(x)\right)^T \in \mathbb{R}^m$ a vector of linearly independent functions that show strong oscillations or even irregular rapid variations. We shall furthermore assume that the functions $\left\{ w_i\right\}_{i=1,\ldots,m}$ satisfy the differential equation system

\begin{equation}
\vec{w}'(x) = A(x) \vec{w}(x)\,,
 \label{lltb:levin:2}
\end{equation}

with an $m\times m$ matrix $A(x)$ containing entries that are varying slowly. The principle of the Levin collocation method relies on finding a function vector $\vec{p}(x) = \left( p_1(x), \ldots, p_m(x)\right)^T$ (or a least an approximation for it) such that $\langle \vec{p}, \vec{w} \rangle' \approx \langle \vec{f}, \vec{w} \rangle$. The integral can then readily be solved:

\begin{equation}
I = \int_a^b{ \langle \vec{f},\vec{w} \rangle \d x} \approx \int_a^b{ \langle \vec{p},\vec{w} \rangle \d x} = \vec{p}^{\,T}(b) \vec{w}(b) - \vec{p}^{\,T}(a)\vec{w}(a)\,.
 \label{lltb:levin:3}
\end{equation}

The problem of evaluating Eq. (\ref{lltb:levin:1}) is therefore replaced by approximating the function vector $\vec{p}$ appropriately. Explicit calculation using the linear independence of the functions $\{w_i\}_{1<i<m}$ yields

\begin{equation}
 \vec{p}\,' + A^T \vec{p} = \vec{f}\,.
 \label{lltb:levin:5}
\end{equation}

Hence, $\vec{p}$ is given by an approximate solution to the ordinary differential equation system. The numerical treatment of this system is feasible, since, by assumption, neither $\vec{f}$ nor $A$ contain rapidly oscillatory components. A solution to the system can, for example, be found by polynomial collocation. 

The function vector $\vec{p}$ is then approximated by a linear combination of basis $n$ polynomials  

\begin{equation*}
\left\{u_k^{(i)} \right\}_{i=1, \ldots, m}^{k=1, \ldots, n} 
\end{equation*}

of degree $n$. An $n$-point approximation $p_i^{(n)}$ of the vector component $p_i$ can be expressed as

\begin{equation}
 p^{(n)}_i(x) = \sum_{k=1}^n {c_k^{(i)} u^{(i)}_k(x)}\,. 
 \label{lltb:levin:6}
 \end{equation}

The coefficients $c_k^{(i)}$ have to be determined by a linear equation system defined by the collocation conditions

\begin{equation}
 \left( \partial_x + A^T\right) p^{(n)}(x_j) = f(x_j) 
 \label{lltb:levin:7}
\end{equation}

at properly chosen collocation points $\left\{ x_j\right\}_{j=1, \ldots, n}$. The corresponding approximation of the integral is then 

\begin{equation}
I \approx I^{(n)} = \sum_{i=1}^m p_i^{(n)}(b)w_i(b) - p_i^{(n)}(a)w_i(a)\,. 
 \label{lltb:levin:8}
\end{equation}

In case of products of two spherical Bessel functions with different arguments (as they appear in Eq. (\ref{lltb:initial:9})), a closed equation system in the form of Eq. (\ref{lltb:initial:11}) can be obtained by considering a four-component function vector given by

\begin{equation}
 \vec{w} = \left( \begin{array}{r}
                   j_\ell(kf(r_i)) \ j_\ell(kf(r_j)) \\
                   j_{\ell-1}(kf(r_i)) \ j_\ell(kf(r_j)) \\
                   j_\ell(kf(r_i)) \ j_{\ell-1}(kf(r_j)) \\
                   j_{\ell-1}(kf(r_i)) \ j_{\ell-1}(kf(r_j))
                  \end{array}
 \right)\,.
 \label{lltb:levin:9}
 \end{equation}

In fact, regarding the recursive expressions of derivatives of spherical Bessel functions

\begin{align}
 \label{lltb:levin:10}
j_\ell'(x) &= j_{\ell-1}(x) - \frac{\ell+1}{x} j_\ell(x)\,, \\
 \label{lltb:levin:11}
j_{\ell-1}'(x) &= -j_\ell(x) + \frac{\ell-1}{x} j_{\ell-1}(x)\,,
\end{align}

this yields a $4\times 4$ matrix of the form

\begin{equation}
A = \left( \begin{array}{cccc}
            -2\dfrac{\ell+1}{k} & f(r_i) & f(r_j) & 0 \\
            - f(r_i) & -\dfrac{2}{k} & 0 & f(r_j) \\
            - f(r_j) & 0 & -\dfrac{2}{k} & f(r_i) \\
            0 & -f(r_j) & - f(r_i) & 2 \dfrac{\ell-1}{k}
           \end{array}
\right)\,,
 \label{lltb:levin:12}
\end{equation}

that fixes the differential equation system of Eq. (\ref{lltb:levin:2}). \bigskip

The Levin collocation method is restricted to definite integrals with finite boundaries which is not the case in Eq. (\ref{lltb:initial:15}). Nonetheless, we perform a change of variable $k \longrightarrow k f(r_i) \equiv \tilde{k} $ and approximate the integral by the expression 

\begin{equation}
 C^\ell(r_i, r_j) =  \frac{2}{\pi f^3(r_i)} \int_{\tilde{k}_\mathrm{min}}^{\tilde{k}_\mathrm{max}} \d \tilde{k} \, \tilde{k}^2 \, P_\Psi(\tilde{k})\, j_\ell( \tilde{k} ) \, j_\ell\left( \tilde{k} \, \frac{f(r_j)}{f(r_i)}\right)\,,
 \label{lltb:levin:13}
 \end{equation}

with suitably chosen boundaries $\tilde{k}_\mathrm{min}$ and $\tilde{k}_\mathrm{max}$. The lower boundary is taken at the $\ell$-dependent ``point of growth" of the spherical Bessel functions (see \cite{tram_computation_2013}) at which the first significant values larger than $10^{-10}$ are obtained. Due to the steep decay of $\tilde{k}^{-7}$ of the integrand in this regime, the integral quickly converges such that the cutoff $\tilde{k}_\mathrm{max}=10^6$ can safely be applied.

 \acknowledgments
 We wnat to thank Matthias Redlich for extensive discussions and for providing a very flexible implementation of the background $\Lambda$LTB model that, in parts, was entering into this work. We thank Bj\"{o}rn-Malte-Sch\"{a}fer for very helpful discussions particularly on the construction of initial conditions. In addition, we are very grateful to Britta Zieser for pointing out the Levin collocation method and an efficient code that could easily be adapted of our purposes. We want to thank the research group of Peter Bastian and their initial help with the DUNE framework. Computations have been performed on the \texttt{bwunicluster} supported by the state of Baden-W\"{u}rttemberg through bwHPC project. We furthermore acknowledge financial support by the German \emph{Deut\-sche For\-schungs\-ge\-mein\-schaft, DFG\/} project number BA 1369 / 20-2. 
 

\bibliographystyle{JHEP}
\bibliography{references/refs}

\end{document}